\documentclass[fleqn,usenatbib]{mnras}

\usepackage{newtxtext,newtxmath}

\usepackage[T1]{fontenc}

\DeclareRobustCommand{\VAN}[3]{#2}
\let\VANthebibliography\thebibliography
\def\thebibliography{\DeclareRobustCommand{\VAN}[3]{##3}\VANthebibliography}

\usepackage{graphicx}	
\usepackage{amsmath}	
\usepackage{xcolor}

\definecolor{darkred}{rgb}{0.76, 0.23, 0.13}

\title[ML approach to photometric metallicities]{A machine learning approach to photometric metallicities of giant stars}

\author[C. P. Fallows and J. L. Sanders]{
Connor P. Fallows\thanks{connor.fallows.20@ucl.ac.uk}
and Jason L. Sanders
\\
University College London, Gower St., London WC1E 6BT, UK\\
}

\date{Accepted XXX. Received YYY; in original form ZZZ}

\pubyear{2022}

\begin{document}
\label{firstpage}
\pagerange{\pageref{firstpage}--\pageref{lastpage}}
\maketitle

\begin{abstract}
Despite the advances provided by large-scale photometric surveys, stellar features -- such as metallicity -- generally remain limited to spectroscopic observations often of bright, nearby low-extinction stars. 
To rectify this, we present a neural network approach for estimating the metallicities and distances of red giant stars with $8$-band photometry and parallaxes from Gaia EDR3 and the  2MASS and WISE surveys. The algorithm accounts for uncertainties in the predictions arising from the range of possible outputs at each input and from the range of models compatible with the training set (through drop-out). A two-stage procedure is adopted where an initial network to estimate photo-astrometric parallaxes is trained using a large sample of noisy parallax data from Gaia EDR3 and then a secondary network is trained using spectroscopic metallicities from the APOGEE and LAMOST surveys and an augmented feature space utilising the first-stage parallax estimates. The algorithm produces metallicity predictions with an average uncertainty of 
$\pm0.19\,\mathrm{dex}$.
The methodology is applied to stars within the Galactic bar/bulge with particular focus on a sample of $1.69$ million objects with Gaia radial velocities. We demonstrate the use and validity of our approach by inspecting both spatial and kinematic gradients with metallicity in the Galactic bar/bulge recovering previous results on the vertical metallicity gradient ($-0.528 \pm 0.002$ dex/kpc) and the vertex deviation of the bar ($-21.29\pm2.74\,\mathrm{deg}$).
\end{abstract}



\section{Introduction}

\begin{figure*}
    \centering
    \includegraphics[width=\textwidth]{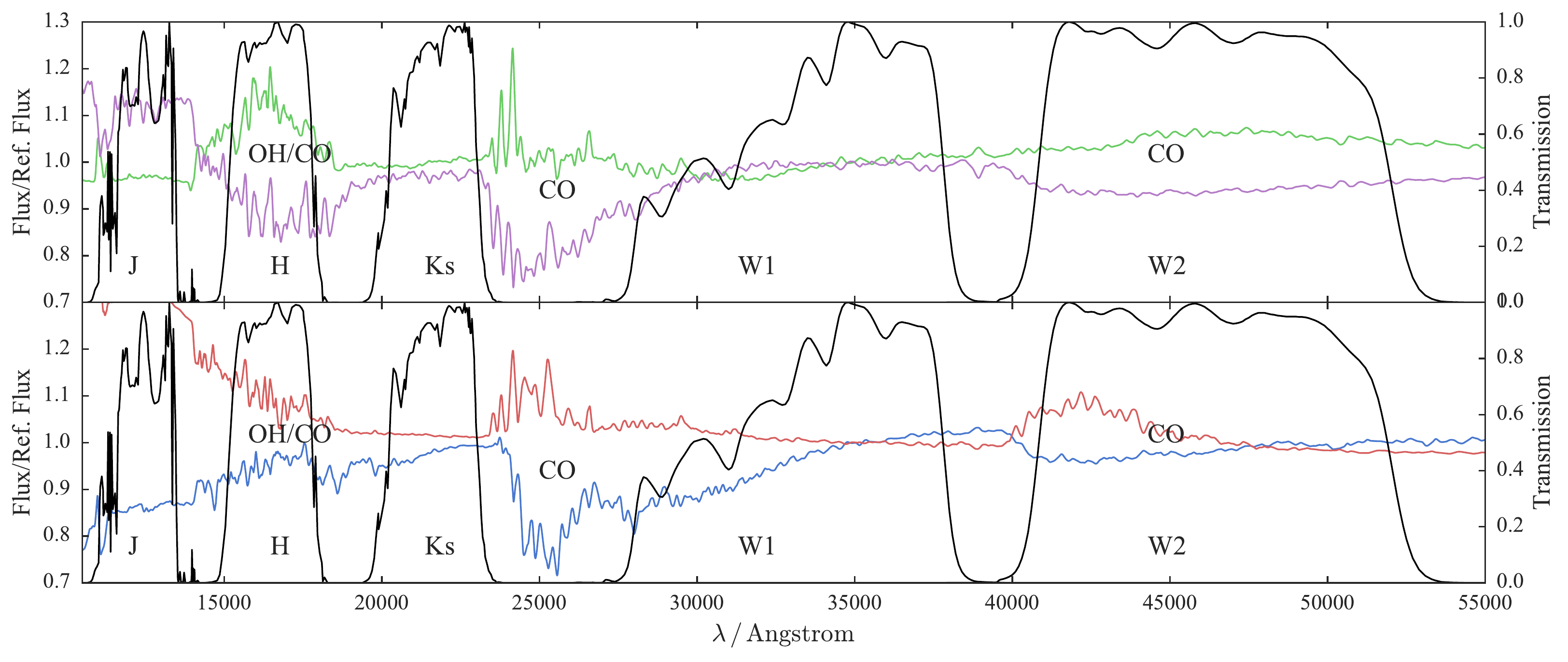}
    \caption
    {The sensitivity of WISE to stellar metallicity: ratios of MARCS models with infra-red bandpasses overplotted normalized at $\lambda=35000\,\mathrm{Angstrom}$. 
    The reference model has $T_\mathrm{eff}=3500\,\mathrm{K}$, $\log g=0\,\mathrm{dex}$ and $[\mathrm{M}/\mathrm{H}]=0\,\mathrm{dex}$. 
    The green and purple lines show models with $[\mathrm{M}/\mathrm{H}]=-1\,\mathrm{dex}$ and $[\mathrm{M}/\mathrm{H}]=1\,\mathrm{dex}$, 
    and blue and red lines $T_\mathrm{eff}=3200\,\mathrm{K}$ and $T_\mathrm{eff}=4000\,\mathrm{K}$. Note the strong gradients in $W2$ due to the CO feature. The effects of temperature and metallicity variations in $W2$ can be distinguished using the bluer 2MASS bands.
    }
    \label{fig:stellar_models}
\end{figure*}
One overarching goal of studying the Milky Way is to reveal its detailed formation and evolution, and place our Galaxy in the context of galaxy formation across the Universe \citep{BHG,Barbuy2018}. With the advent of large-scale spectroscopic surveys (RAVE, \citealt{Steinmetz2020}, APOGEE, \citealt{Ahumada2020}, LAMOST, \citealt{Cui2012}, Gaia-ESO, \citealt{GaiaESO}, SEGUE, \citealt{Yanny2009}, GALAH, \citealt{Buder2020}, and in future DESI, \citealt{DESI}, WEAVE, \citealt{WEAVE}, 4-MOST, \citealt{Jong21019}, Milky Way Mapper, \citealt{MWM}), we have highly detailed observations of $>10^6$ stars allowing characterisation of their effective temperatures, surface gravities, radial velocities, chemical compositions, masses, ages and more, from which we can make progress on this goal by elucidating and separating the series of events and processes that have shaped our Galaxy over cosmic time.

However, despite the utility of spectroscopic data, these surveys do have limitations of scope when applied to some problems. As noted by \citet{Ivezi2008} and \citet{Huang2021}, taking spectroscopic data for very distant or faint objects can quickly become difficult. This causes many surveys to have complex selection criteria to ensure good spectroscopic data can be taken. These criteria typically limit observations to specific object classes within a limited sky region making the application of such data to large-scale populations or structures difficult, as only a small portion of these groupings may be included in the selection criteria. For example, when attempting to study the inner regions of the Milky Way's disk and bulge, the large distances and extreme extinction effects put many stars beyond the reaches of spectroscopic observations. This area of the sky therefore tends to have relatively few spectroscopic observations, which makes analysis of these interesting populations difficult (although the infra-red APOGEE and in future MOONS, \citealt{MOONS}, surveys are rapidly changing this state of affairs).

On the other hand, we have large-scale photometric surveys, which typically are not bound by the same criteria that tend to limit spectroscopic observations. This allows them to be far more expansive, generally observing many classes of object across the whole sky (or often at least half) to a significantly greater depth. For example, Gaia \citep{Gaia2021}, 2MASS \citep{Skrutskie2006} and WISE \citep{Wright2010} have all observed the entire sky across the optical to infrared, whilst SDSS \citep{SDSS}, Pan-STARRS \citep{PanSTARRS}, DES \citep{DES}, Sky-Mapper \citep{SkyMapper} and GALEX \citep{GALEX} among others have surveyed large fractions of the sky. However, unless designed with filters with specific sensitivity to stellar metallicity or surface gravity like Sky-Mapper's \emph{u} and \emph{v} bands \citep{SkyMapper_design}, broad-band photometric data tends to struggle to accurately determine stellar parameters without additional input.

Thus, we reach our aim with this research: to develop a method that can determine stellar properties with the utility of spectroscopic data, while retaining the scope and scale of photometric surveys. From this, we would then be able to analyse, on a much deeper level, the stellar populations and structures that stretch across the Milky Way.

Attempts to determine stellar metallicity from photometry have had some past successes, through leveraging the subtle sensitivity of broad-band colours to metallicity. 
One early approach was the `UV excess' method \citep{Wallerstein1962} which can be calibrated to map both stellar temperatures and metallicities using the large number of metal lines in bluer and ultraviolet bandpasses. This method has been adapted for use with modern photometric surveys, using the SDSS \citep{Ivezi2008} and Pan-STARRS \citep{Thomas2019} $(g - r)$ and $(u - g)$ colours to estimate metallicities. In a similar vein, the metallicity sensitivity of the Ca H\&K region at $\sim3950$\AA\ has been targeted using narrow-band filters in the PRISTINE \citep{PRISTINE} and Sky-Mapper \citep{SkyMapper} surveys \citep[see][for catalogues of stellar parameters derived from Sky-Mapper data]{Huang2021,Lin2022}. Due to the strong effects of extinction on UV/near-UV, these methods are less effective for studying faint or distant objects within the highly-extincted inner Milky Way (although see \citealt{PIGS} for a study of metal-poor stars in the Galactic bulge using the PRISTINE survey).

For more highly extincted regions, infra-red photometric surveys are more attractive. \cite{Schlaufman2014}, \cite{Koposov2015}, \cite{Li2016} and \cite{Casey2018} have all demonstrated how the infra-red WISE survey  \citep{Wright2010} can be used to both separate dwarf and giant stars and also estimate stellar metallicities for red stars. In particular, the WISE colour $(W1-W2)$ displays a strong correlation with stellar metallicity ($W1$ and $W2$ have effective wavelengths of $3.4\,\mu$m and $4.6\,\mu$m respectively).
This is primarily due to the presence of a CO feature in the spectrum of M giants.
In Fig.~\ref{fig:stellar_models} we show ratios of stellar spectra from the MARCS model grid \citep{MARCS}. 
Increasing the metallicity we observe the molecular features (particularly the CO band in the $W2$ bandpass) weaken whilst the flux in $K_s$ and $W1$ are essentially unaffected leading to bluer $(K_s-W2)$ and $(W1-W2)$ for more metal-rich stars. These colours also vary with effective temperature (redder for hotter stars) again due to CO variation but this degeneracy can be removed by combining with bluer colours such as $(J-K_s)$.
This metallicity sensitivity of the WISE bands was utilised most recently by \citet{Grady2021}, who used machine-learning regression models with Gaia, 2MASS, and WISE bands to estimate metallicities of stars in the Magellanic Clouds. This improved on past works by allowing the subtler metallicity sensitivity of other photometric colours to be included. For example, they found that by including Gaia $(G_\mathrm{BP}-G_\mathrm{RP})$ and 2MASS $(J - H)$ they were able to add additional metallicity information beyond that provided by $(W1 - W2)$. This work provided metallicity estimations with high accuracy ($\pm0.13$ dex for $-1\leq\mathrm{[Fe/H]}\leq-0.5$) allowing for detailed mapping of the mean metallicity of the Magellanic Clouds. However this method was not used on stars within the Milky Way.

Past research has therefore left a gap for broadly applying metallicity estimation to large-scale photometric surveys of the Milky Way. It should be noted that WISE information is often utilised in stellar characterization pipelines that provide metallicity estimates \citep[e.g.][]{Anders2022,Lin2022} although these methods rely on theoretical stellar models, or isochrones, which can be uncertain for cool stars with significant molecular contributions to their atmospheres. Here we provide a complementary data-driven approach to instead learn the correlations between photometric colours and  metallicities obtained from large spectroscopic surveys. We thus bypass complexities in detailed stellar modelling.
In doing this, we supplement Gaia EDR3 \citep{Gaia2021} astrometry with metallicity information. Such a combination allows us to study the spatial, kinematic, and abundance trends within the Galaxy, and, thus, we are able use this new methodology to probe the evolution and origins of various Milky Way structures. 

This paper is split into four main components: neural-network setup, distance estimation, metallicity estimation and a brief analysis of the properties of a bar-bulge sample. Section~\ref{NNsetup} describes the general setup of the neural network algorithm we will use in the subsequent methods. Section~\ref{Distance_Est} describes our machine-learning-enhanced approach to refine the distances we use in our analysis, allowing us to improve object positional information and refine absolute magnitude calculations. Section~\ref{Feh_Est} covers the estimation of metallicities through the use of our neural-network algorithms, and the creation of our final output catalogue. Section~\ref{section::results} describes an investigation of the spatial and kinematic gradients of a bar-bulge sample separated by our photometric metallicities, before we close with our conclusions in Section~\ref{section::conclusions}.

\section{Neural Network Setup}
\label{NNsetup}
For the most accurate predictions of photometric metallicity, we opt for a neural network (NN) machine learning algorithm. Typically, NN architectures are trained with a set of input features, which are fed through a non-linear layered network to return an output value. The network layers are constructed from a set of inter-connected nodes, with the strength of the connections (or weights) tuned through training to allow the model to learn patterns in the input data. Training is guided by the network's `loss function', which guides the penalty the model receives for returning poor predictions of the outputs compared to the training set, and which the network aims to minimise. The most common loss function is the mean squared error between the NN's output and desired target values -- although this can be customised and tuned for the desired setup. 

Using the Python implementation of the Torch machine learning library, Pytorch \citep{Paszke2019}, we work with a NN with the architecture shown in Fig.~\ref{fig:fig_one} and described in-detail in Appendix~\ref{appendix_NNsetup}. For a set of input features, $x$, each layer in the network, $h$, follows 
\begin{equation}
    h = a f(x) + b,
	\label{eq:h_eq}
\end{equation}
where $f(x)$ is the non-linear response function of the layer, and the matrix of weights $a$ and vector of biases $b$ are constants refined by the training process. Thus, for a network of $n$ layers, we return an output value, $y$, from outputs of one layer being sequentially input to the next. This gives us:
\begin{equation}
    y = a_{n} f( a_{n-1} f(\dots f(a_2f(a_{1} f(x) + b_{1}))+b_2\dots) + b_{n-1} ) + b_{n}.
	\label{eq:h_repeat_eq}
\end{equation}
The network is trained iteratively with the network tuning $a_i$ and $b_i$ to improve the loss function. By improving the average loss function over the full training set, the network is able to learn the correlation between $x$ and $y$ and predict outputs for new sets of input features. This allows for accurate and robust fitting, while also allowing $f(x)$ to be customised and modified to best suit a chosen problem.

However, NN's do not tend have a measure of `confidence' in their estimations and instead usually return a single value for a set of input features. For us to include a measure of the network's predictive confidence, we add two small modifications adapted from \citet{Leung2019a}: an uncertainty output node and node drop-out.

\subsection{Uncertainty Node}

\begin{figure}
	\includegraphics[width=\columnwidth]{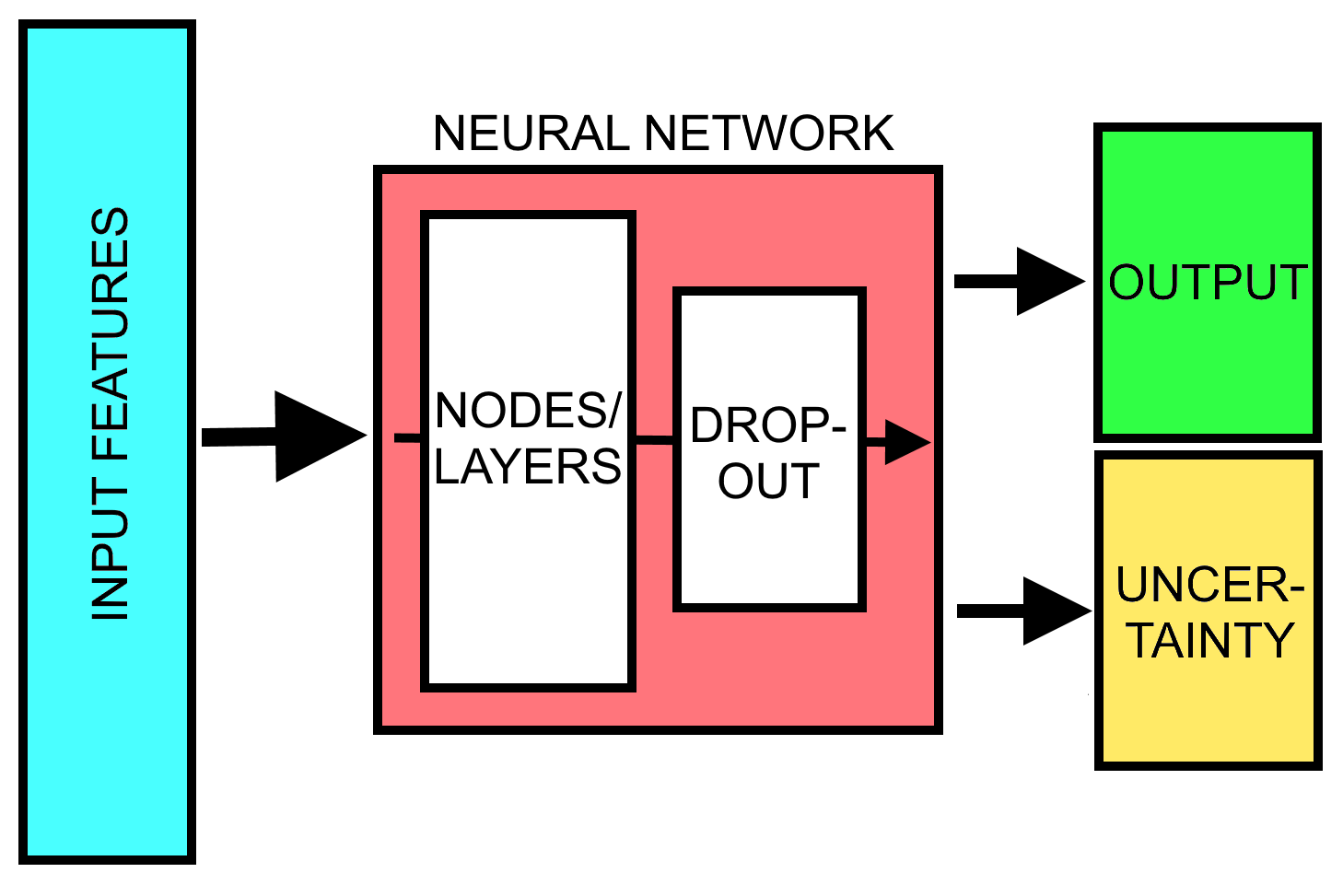}
    \caption{Diagram of the adopted Neural Network (NN) architecture. Input features are fed into the NN, which trains the nodes/layers with the drop-out modifier active. Then, for predictions, the layers predict a value with a combined uncertainty from the drop-out stochasticity and the secondary `uncertainty' output node.}
    \label{fig:fig_one}
\end{figure}

We include a secondary output node into the NN architecture, as marked in Fig.~\ref{fig:fig_one}. This node provides one uncertainty measure, $\sigma_{\mathrm{pred}}$, known as the model's `predictive uncertainty'. This is the variance in the training data that isn't accounted for by the uncertainties noted in the training set's output targets. Even with perfect data, there are `hidden variables' that impact the outputs. This manifests as identical training inputs into the NN returning a range of outputs. During the training process, the output uncertainty was fed into the NN’s customised loss function, and allows us to return an output value along with an uncertainty measure. 

We adopt the loss function from \citet{Leung2019a}. With $y_{i}$ as the target value from the training set with uncertainty $\sigma_{\mathrm{data}, i}$, and $\hat{y}_{i}$ the value returned by the NN with uncertainty $\sigma_{\mathrm{pred}, i}$ (from the `uncertainty node'), the logarithm of the joint variance is determined as $s_{i} = \ln(\sigma^{2}_{\mathrm{data}, i} + \sigma^{2}_{\mathrm{pred}, i})$. The loss function, $J(y_{i}, \hat{y}_{i})$, is then defined as
\begin{equation}
    J(y_{i}, \hat{y}_{i}) = \frac{1}{n} \sum_{i=1}^{n} \frac{1}{2} (y_{i} - \hat{y}_{i})^2 e^{-s_{i}} + \frac{1}{2} s_{i}. 
	\label{eq:one}
\end{equation}
The predictive uncertainty, $\sigma_\mathrm{pred}$ is refined by the training process, with each iteration of training incrementally refining the uncertainty output when calculating the loss function. The function in equation~\eqref{eq:one} is designed such that the network minimises loss from poor predictions by maximising the predictive uncertainty. However, this drive is countered by the final additive term which increases loss for high predictive uncertainty. In this way, the network optimises to find the largest predictive uncertainty for the given data, but is penalised for selecting extremely large or small values.

\subsection{Drop-Out}

Drop-out \citep{Hinton2012} is a common NN operation used to dissuade over-fitting during the training stage by randomly `dropping’ a fraction of the nodes in each layer. This modifies equation~\eqref{eq:h_eq} to be
\begin{equation}
    h = a \; g(x) + b,
	\label{eq:h_eq_gx}
\end{equation}
where $g(x) = \mathrm{P} \, f(x)$. Here, P is a function that applies a Bernoulli distribution to each node within a layer (and thus modifies the response of $f(x)$). The Bernoulli function causes some chosen fraction of nodes within a layer to be temporarily `zeroed' out, and thus have no effect on the current training or prediction pass. This limits the effect one node or branch can have to the overall output, as other nodes in the network must learn to `cover’ for those hidden by the drop-out process. With drop-out active, the network tends to learn the problem as a cohesive unit, and avoids the creation a small number of over-influential nodes that can dictate the network's predictions. 

However, in our case, drop-out can have a secondary function to add stochasticity to the model \citep{Gal2015}. As each run of the network has a random fraction of nodes missing, we can consider each run to be a slightly different network. So, if a set of input features are repeatedly passed through the NN, the variations due to drop-out will cause different outputs to be returned each time. While predicting from our network, we return an ensemble of networks with slight variations due to the randomness of drop-out, all of which are consistent with the training data. When we input new features, we return a distribution of output values from the ensemble of networks - a distribution we can consider as a probability. We therefore consider our model to be a Bayesian Neural Network, which returns a probabilistic distribution rather than a single value. From such a distribution, we draw a prediction (mean) and an implicit uncertainty (standard deviation). 

We return our uncertainty from the drop-out stochasticity as $\sigma_{\mathrm{drop}}$. Then, with both the drop-out uncertainty, $\sigma_\mathrm{drop}$, and the predictive uncertainty, $\sigma_\mathrm{pred}$, calculated, we can determine the final uncertainty of each prediction, $\sigma_{\mathrm{total}}$, by

\begin{equation}
    \sigma_{\mathrm{total}} = \sqrt{\sigma_{\mathrm{drop}}^{2} + \sigma_{\mathrm{pred}}^{2}}. 
	\label{eq:two}
\end{equation}
\section{Distance Estimation}
\label{Distance_Est}

Before we begin estimating metallicities, our method requires a robust measure of stellar distances. Distances allow us to calculate absolute magnitudes for our sample stars, which can provide essential information on intrinsic stellar properties for the NN's model. 

While using Gaia parallaxes directly would be the ideal choice for data-driven analysis, there are a number of limitations to such an approach. As described by \citet{BailerJones2021}, transforming between parallax and distance can lead to issues if done naively. Objects with $\varpi \approx 0$, even with well-constrained uncertainties, will tend to have very large fractional errors. This equates to extremely large distance uncertainties for stars beyond a few kiloparsecs. Additionally, valid parallaxes in the Gaia catalogue can have negative values due to the random scatter from uncertainties at small parallaxes, which makes the naive $r = 1/\varpi$ relation impractical to apply. The approach developed by \citet{BailerJones2015} instead adds a statistical prior to distance prediction that works to guide estimates for objects with poorly informative Gaia parallaxes. From this, a distance estimation can be drawn allowing us to avoid the limitations of the raw Gaia data.

However, for this work, we aimed to focus on a large proportion of the Milky Way's stars. Therefore, many of the objects in our samples exist in the distance regime where prior information becomes dominant over Gaia parallax information. While this was expected behaviour for this approach, we found some estimates to be strongly dependent on the parameters of the prior rather than being guided by Gaia measurements -- which reduce the utility of these distances for our methodology.

In an attempt to reduce the impact of the prior, \cite{BailerJones2021} adjust their method to also include a star’s photometric information (producing `photogeometric' distances). Briefly, this secondary approach uses a colour-magnitude prior (derived from Gaia photometric bands) to restrict the range of absolute magnitudes an object of a given colour can have. Thus, they constrain the distance probability function their method returns. With this addition, they find an improvement in the precision of stars with poorly informative parallaxes.

Our approach follows on from this idea, expanding the addition of photometric information through the inclusion of a wide range of additional bands \citep[see][for a similar approach also utilising spectroscopic information]{Hogg2019}. However, instead of using this data as a constraint on our distance estimates, we instead used our photometric information and NN algorithm to estimate an independent parallax value. This `photometric parallax' was then combined with the parallaxes from Gaia, and allowed us to return values with much lower uncertainties. Thus, we reduced the regime where parallax information is uninformative, and thus limited the number of objects where the \cite{BailerJones2021} prior has a significant impact.

\subsection{Data Collection}
\label{Dist_data_coll}

In order to augment existing distance information with photometric data, we required accurate astrometry and a wide range of photometric colours.

We followed the lead of \citet{Grady2021}, and selected our data from three photometric surveys: Gaia EDR3 \citep{Gaia2016, Gaia2021, Riello2021, Seabroke2021}, 2MASS \citep{Skrutskie2006}, and the unWISE catalogue \citep{Schlafly2019}. The Gaia survey is an optical photometric survey, with three bands (G, G$_{\mathrm{RP}}$, \& G$_{\mathrm{BP}}$) between 330nm and 1050nm , and focusses on observing accurate sky positions, proper motions, parallaxes, and radial velocity information. The 2MASS survey instead observes in near-infrared, with three bands, J, H, and Ks, with peak sensitivity at 1235nm, 1662nm, and 2159nm respectively, which grants information to separate giant and main-sequence stars \citep{Majewski2003} as well as bolster extinction measurements (as will be discussed later). Finally, the unWISE survey is built upon the results of the WISE catalogue described previously \citep{Wright2010}, but with altered image processing to retain observation resolution in star-dense regions. This increases the available number of objects with WISE bands (W1, W2, W3, W4 at 3.4$\mu$m, 4.6$\mu$m, 12$\mu$m, and 22$\mu$m respectively), and thus greater coverage at large distances and within high-density sky regions. For our sample, we avoided the W3 and W4 bands due to the small number of objects with accurate observations, which would have limited the maximum potential size of our sample.

With access to the H and W2 bands, we made use of the Rayleigh-Jeans Colour Excess (RJCE) Method \citep{Majewski2011} to determine accurate extinction corrections for objects in our sample. This approach relies on the fact that, for most stellar types, intrinsic (H-W2) colour is nearly constant. Therefore, significant reddening in this colour can provide a good measure of the extinction effects on a star-by-star basis. To transform the extinction to the other photometric bands, we used the extinction coefficients from \cite{Wang2019}.

Objects were chosen to ensure good photometry by filtering for high-quality observations. We limit the Gaia BP/RP flux excess to $\leq3.0$, limit the astrometric renormalised unit weight error to values $\leq1.4$, and select only for objects with `good' W1 \& W2 photometry from the UnWISE quality flags. We further ensured all of our sample had velocity information (proper motion, radial velocity) from Gaia, which provided kinematic information for stars within our sample. This kinematic information, when combined with the distances from our method, could then allow us to calculate three-dimensional velocities for each star, and thus analyse the kinematic distributions of our sample objects. Due to the limitations of Gaia’s radial velocity measurements, requiring radial velocities remained the largest limit on our sample size, with only 0.4\% of the full survey catalogue having radial velocity information. Corrections to Gaia parallaxes were also made at this stage, accounting for zero-point errors in the measurements described in \citet{Lindegren2021}.

\subsection{Methodology}
\label{section:Method}
Our method leveraged the NN architecture described in Section~\ref{NNsetup}, and a `pseudo-absolute magnitude' measure described by \citet{Arenou1999}. We anticipated that, if we chose to have the NN predict a value of parallax directly from photometric data, the algorithm would struggle to learn a direct correlation between a star’s photometric colour and its distance. However, as the absolute magnitude of a star is intrinsic to the star, it is therefore independent of object distance. We were thus able to use this as a target for the NN to predict, rather than attempting to estimate parallax directly.

We used the pseudo-absolute magnitude defined in the 2MASS J-band, $M_{J, \mathrm{pseudo}}$, as the basis for our analysis, where $\varpi$ was the Gaia parallax and $J_{c}$ was the extinction-corrected apparent J-band magnitude. $M_{J, \mathrm{pseudo}}$ was therefore defined as
\begin{equation}
    M_{J, \mathrm{pseudo}} = \varpi 10^{0.2J_{c}}.
	\label{eq:Mjp}
\end{equation}
This value acted as a good proxy for absolute magnitude by combining parallax and magnitude information. The NN therefore made it's predictions within the pseudo-absolute magnitude parameter space, rather than parallax space, and was therefore generalisable beyond the scope of the training data. Had we estimated parallax alone, the NN would struggle to predict reliably towards (and beyond) the edges of the parameter space --- especially towards distant object parallaxes at the smallest end of our range. Furthermore, this formulation for pseudo-absolute magnitude allowed the Gaussian uncertainties in parallax to be translated into Gaussian uncertainties in pseudo-absolute magnitude space.

\begin{figure}
	\includegraphics[width=\columnwidth, trim={20, 20, 40, 40}]{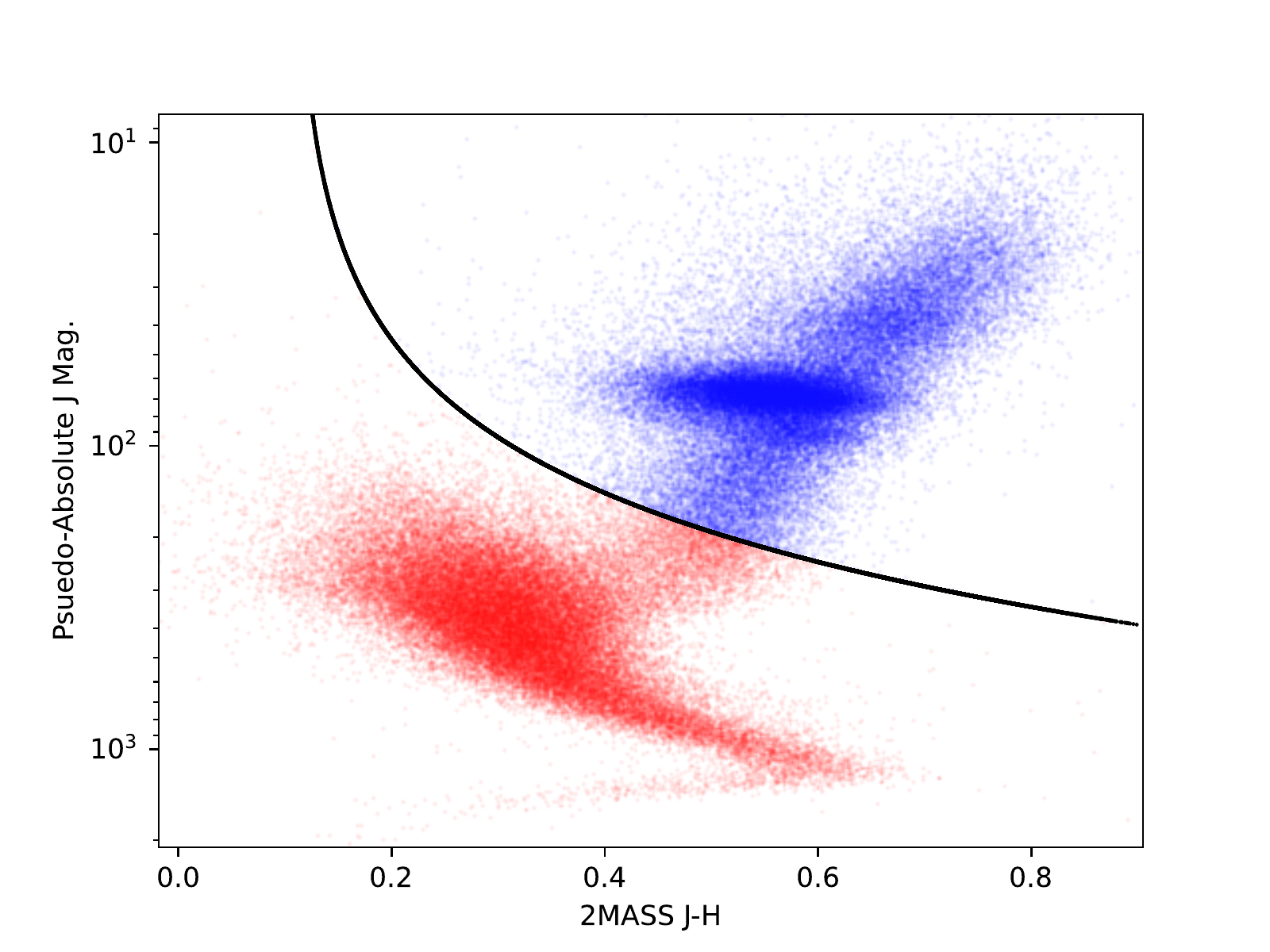}
    \caption{Plot of object pseudo-absolute J magnitudes against J-H colour. Both axes have been corrected for extinction using the RJCE method \citep{Majewski2011}. The giant-dwarf cut is shown with the main-sequence in red, and the red clump/giant branch in blue.}
    \label{fig:aMK_vs_JH}
\end{figure}

Initially, we used this value to filter our sample for giant stars. Due to their intrinsic brightness, giant stars are ideal targets for long distance analysis of the Milky Way's population. Therefore, to avoid the NN's attention being split between dwarf and giant stars while training -- and thus lowering the model's overall performance -- we removed non-giant stars from our sample. The giant and dwarf populations were clearly visible in colour-pseudo-absolute magnitude space, and so we were able to apply a simple cut in these parameters. With $M_\mathrm{J}$ and $M_\mathrm{H}$ being 2MASS J and H extinction-corrected apparent magnitudes respectively, and the J-band pseudo-absolute magnitude being $M_{J, \mathrm{pseudo}}$, we selected only objects where
\begin{equation}
   M_{J, \mathrm{pseudo}} < 492.101(M_{\mathrm{J}} - M_{\mathrm{H}}) - 53.827.
	\label{eq:giant_cut}
\end{equation}
This cut is shown clearly in Fig.~\ref{fig:aMK_vs_JH}, separating the red clump and giant branch from the main sequence.

From this, we set the NN to accept 16 photometric colours as our input array, $x$, (described in Appendix~\ref{appendix_NNsetup}), and to predict the pseudo-absolute magnitude, $y$. 
As mentioned previously, using $M_{J, \mathrm{pseudo}}$ had the notable advantage of inheriting the Gaussian uncertainties of the Gaia parallaxes. We then returned a NN-refined parallax value, $\mu_{\mathrm{NN}}$, as
\begin{equation}
    \mu_{\mathrm{NN}} = M_{J, \mathrm{pseudo}}10^{-0.2J_{c}},
	\label{eq:Mjp_10jc}
\end{equation}
and similar for the uncertainty $\sigma_\mathrm{NN}$ from the uncertainty node.

To train the network, we followed a method of `cross-training' which functioned similarly to common cross-validation methods.

As every object in our data sample has a Gaia parallax, we chose to train the NN on our sample rather than some external source. In order to train our sample, we split our sample into eight equal `chunks' which we iterated through. For each chunk, the remaining $\sim88$\% was used to train our network, and returned new parallaxes for objects within the chosen chunk. Between each iteration, we reset the NN's training for the new chunk, which avoids biases that arising from objects appearing in both the training and prediction datasets.

With the network's predictions applied to our entire sample, we had derived a set of parallaxes from stellar photometry alone. We therefore considered these results as independent measurements to the parallaxes reported by Gaia. Thus, we combine the two values to improve the overall parallax uncertainty. For an object with a Gaia parallax, $\mu_{\mathrm{Gaia}}$, and associated uncertainty, $\sigma_{\mathrm{Gaia}}$, and with a NN-predicted parallax, $\mu_{\mathrm{NN}}$, and associated uncertainty, $\sigma_{\mathrm{NN}}$, we calculate our combined parallax, $\mu_{\mathrm{new}}$, and combined uncertainty, $\sigma_{\mathrm{new}}$ as
\begin{equation}
    \mu_{\mathrm{new}} = \left(\frac{\mu_{\mathrm{Gaia}}}{\sigma_{\mathrm{Gaia}}^{2}} + \frac{\mu_{\mathrm{NN}}}{\sigma_{\mathrm{NN}}^{2}}\right) \left(\frac{1}{\sigma_{\mathrm{Gaia}}^{2}} + \frac{1}{\sigma_{\mathrm{NN}}^{2}}\right)^{-1},
	\label{eq:four}
\end{equation}
and 
\begin{equation} 
    \frac{1}{\sigma_{\mathrm{new}}} = \sqrt{\frac{1}{\sigma_{\mathrm{Gaia}}^{2}} + \frac{1}{\sigma_{\mathrm{NN}}^{2}}}.
	\label{eq:five}
\end{equation}
Therefore, we produced a unified parallax value with much narrower error than the initial Gaia parallaxes, reducing the number of stars with poorly-informative parallaxes --- and so reduced the proportion of objects for which the Bailer-Jones statistical prior was dominant for distance estimation. 

\begin{figure}
	\includegraphics[width=\columnwidth, trim={20, 20, 40, 40}]{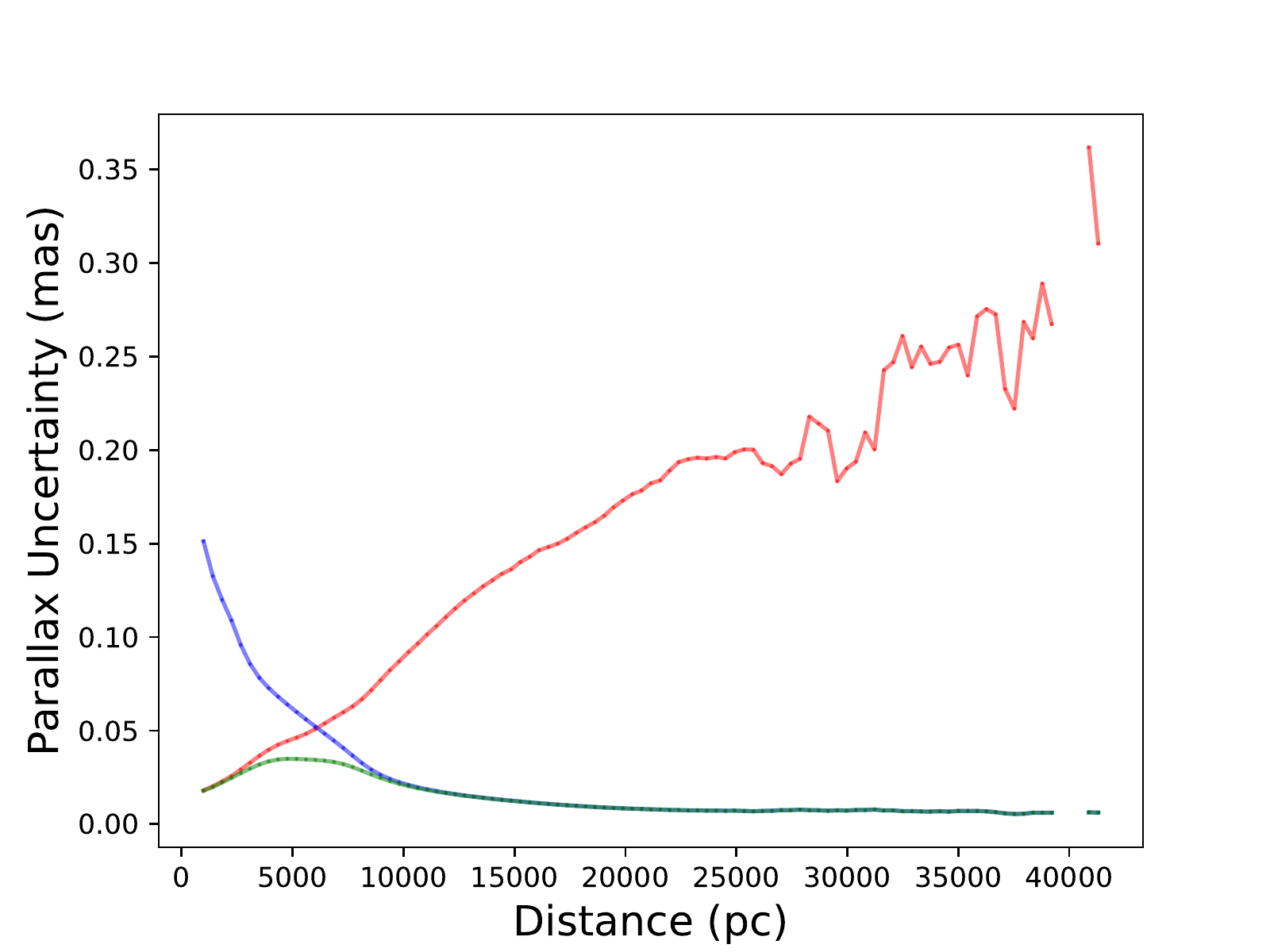}
    \caption{Plot of parallax uncertainty against object distance for Gaia parallaxes (red), our NN parallaxes (blue), and the combined unified parallax (green). Note that this sample is not limited to only objects with Gaia radial velocities.}
    \label{fig:para_over_para_error_rgb}
\end{figure}

We note this improvement in Fig~\ref{fig:para_over_para_error_rgb}, where we show how parallax uncertainties vary with distance for our Gaia data, our NN's outputs, and for the unified parallax value. For this comparison, we have removed the limit of only selecting objects with Gaia radial velocity information. As objects with radial velocities will tend to be brighter, Gaia parallaxes tend to be good, and our NN-based approach has limited impact. Removing this limit shows a more general comparison between the Gaia and NN parallax performance, and highlights clearly where our method provides improvement. It is clear that, at around 6.1kpc, our NN parallax uncertainties become smaller than those from Gaia. At distances beyond this, our parallaxes are therefore more informative than those from Gaia, and our unified value retains a low uncertainty out to larger distances.

With each object given a NN-enhanced parallax, we calculated new distance estimations. We apply the method of \citet{BailerJones2021}, which uses the parallax information and simulation-backed prior distributions to return an estimated parallax. Our NN-enhanced parallaxes form a notable reduction in the number of uninformative parallaxes, decreasing the parallax uncertainty for around 58\% of our overall sample. When we focus only on objects with Gaia parallax/uncertainty < 2.0, we find around 89\% of objects see an improvement from our method. These distances were taken forward to calculate objects positions and absolute magnitudes.

\subsection{Validation}

To validate the accuracy of the distance predictions, we had three measures: the uncertainty output calculated by the NN, and two samples with comparison distance estimates. These comparison distance samples were those calculated by \citet{BailerJones2021}, and those calculated from the AstroNN algorithm \citep{Leung2019b}. 

\subsubsection{Network Uncertainty}

\begin{figure}
	\includegraphics[width=\columnwidth, trim={20, 20, 40, 40}]{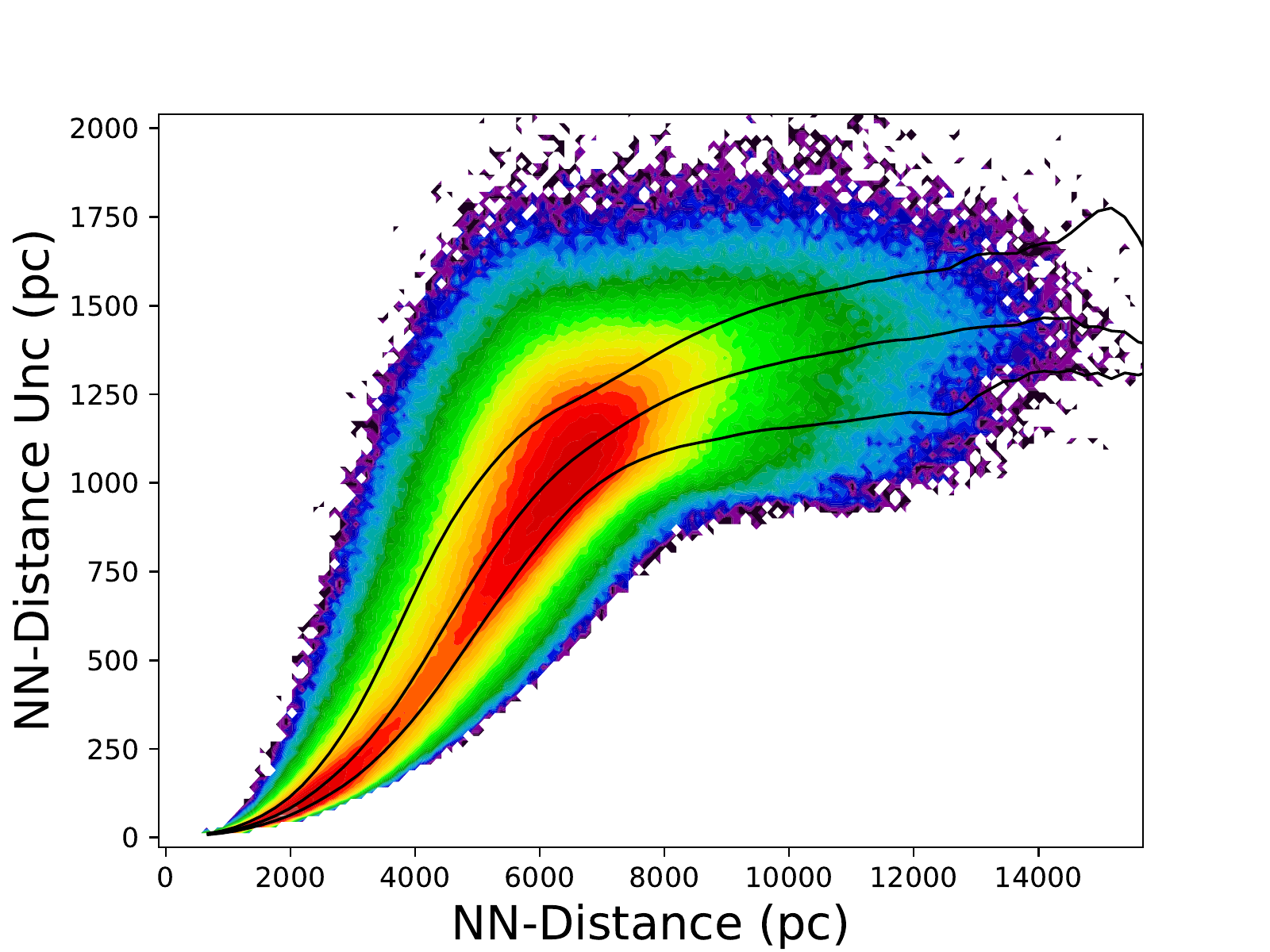}
    \caption{Plot of NN-estimated distance uncertainties against the corresponding absolute distances. Note that these contour plots are logarithmic, and the upper and lower curves are the 84th and 16th percentiles respectively. Further, note that this sample is not limited to only objects with Gaia radial velocities.}
    \label{fig:dist_distu}
\end{figure}

From the NN, we obtained a predicted value of distance (and an associated uncertainty) for each object. We found that this value is low for the majority of our sample, with the mean uncertainty of our whole sample being $\pm159.7\,\mathrm{pc}$.  We plot these uncertainties versus estimated distances in Fig.~\ref{fig:dist_distu}, with uncertainties binned by absolute distance shown in Table~\ref{tab:dist_unc_tab}. We note that, as with Fig.~\ref{fig:para_over_para_error_rgb}, this plot is not limited to only objects with Gaia radial velocity information. As discussed in Section~\ref{section:Method}, this gives us a better sense of our NN's performance than if we only focus on the brighter sample with radial velocity data.

As expected, distance uncertainties remained small for closer objects, and become larger for distant objects. The distance uncertainties remained below 10\% for objects closer than approx. 6kpc, with the the furthest objects in our sample having distance uncertainties less than 20\%.

\begin{table}
    \centering
    \caption{Table of mean distance uncertainties for binned absolute distance ranges. All distances are reported in pc. Percentage uncertainties are taken with respect to the mid-point of the bin.}
    \label{tab:dist_unc_tab}
    \begin{tabular}{ccc}
        \hline
        Distance Bounds & Distance Unc. & Unc. Percentage \\ \hline
        $0 <  d < 2000$  & ± 35.018 &  3.5\%      \\ 
        $2000 < d < 4000$ & ± 110.750 & 3.7\%        \\ 
        $4000 < d < 6000$ & ± 369.182 & 7.38\%        \\ 
        $6000 < d < 8000$ & ± 748.284 & 10.67\%        \\ 
        $8000 < d < 10000$ & ± 1142.063 & 12.69\%        \\ 
        $10000 < d < 12000$ & ± 1571.712 & 14.28\%         \\ 
        $12000 < d < 14000$ & ± 2072.115 & 15.94\%         \\ 
        $14000 < d < 16000$ & ± 2672.450 & 17.81\%         \\ 
        \hline
    \end{tabular}
\end{table}

\subsubsection{\citet{BailerJones2021} Distance Comparison}

\begin{figure*}
	\centering
	\includegraphics[width=.48\textwidth, trim={0, 0, 40, 0}]{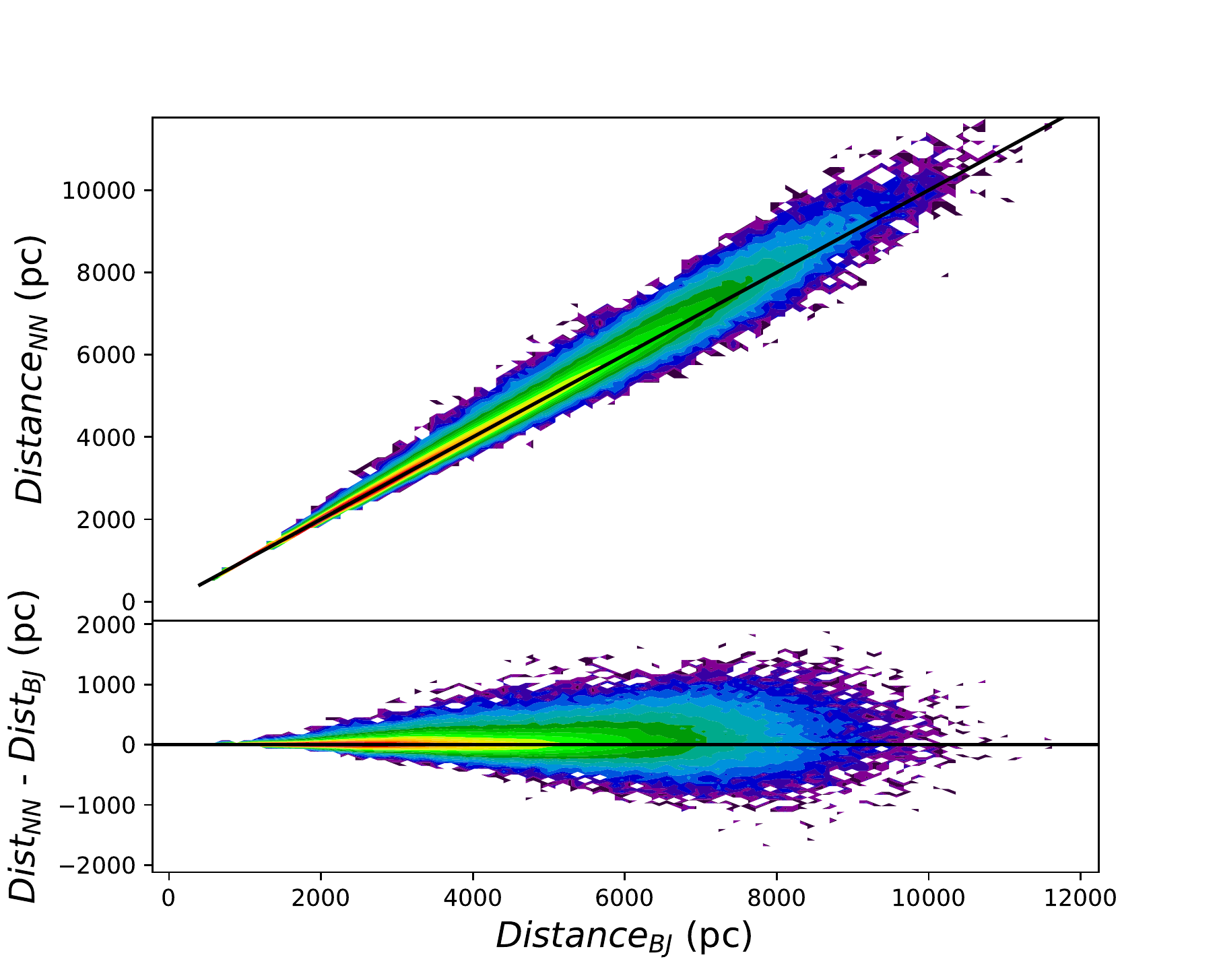}
	\includegraphics[width=.48\textwidth, trim={0, 0, 40, 0}, clip]{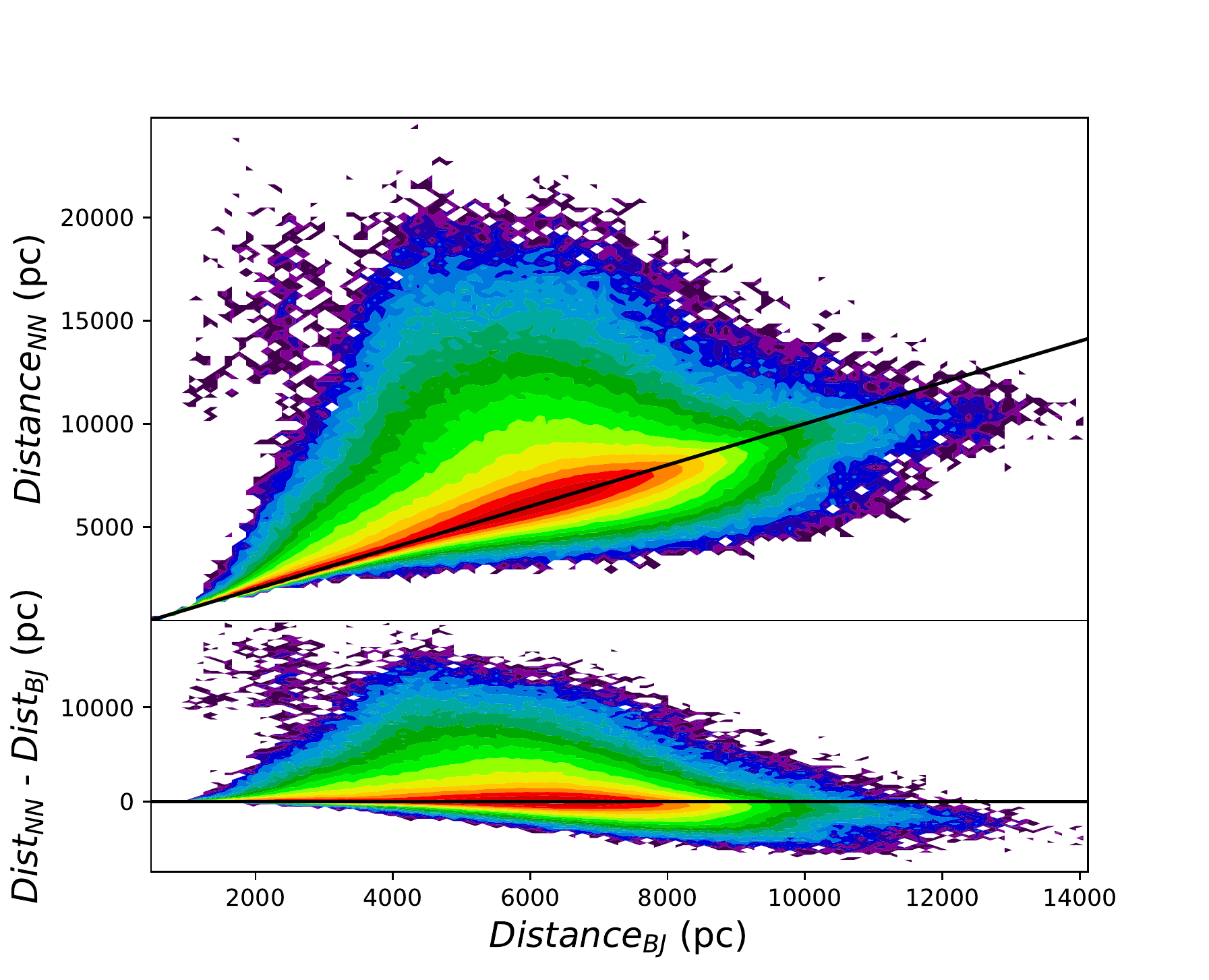}
    \caption{Logarithmic contour plot of NN distances from our work vs. distances from \protect\cite{BailerJones2021} for our main sample (left). We also perform the same comparison for a sample without the prerequisite of radial velocity data allowing comparisons out to greater distance (right).}
    \label{fig:fig_Bjcomp}
\end{figure*}

We compared our network’s performance in comparison to the photogeometric values calculated by \citet{BailerJones2021}. These reference distance values were the values we initially hoped to improve upon with our method. We used much the same method, but applied our NN to reduce the impact of prior terms on the distance estimates. It was therefore expected for there to be good agreement between the two datasets where parallaxes are highly-informative, and significant divergence in the regime where fractional Gaia parallax uncertainties were large (i.e. very high uncertainty, or very small parallaxes) and the NN had a stronger influence. If our approach returned accurate distances, we expected to see the majority of stars match between the two samples, with divergences in parallax space remaining symmetric and becoming more prominent for objects further away. We show this comparison for our sample in the left panel of Fig.~\ref{fig:fig_Bjcomp}, where we see a clear correlation between the two methods (for good parallaxes) with a large scatter due to the impact of the NN. 

We further highlight the right panel of Fig.~\ref{fig:fig_Bjcomp}, where we selected a larger comparison sample without the restriction of requiring radial velocity data for all stars. This allowed us to observe additional objects at large distances ($\geq10\,\mathrm{kpc}$) as well as fainter objects at closer ranges. The minor over-estimation bias for \cite{BailerJones2021} distances between 4 kpc and 8 kpc appeared to be due to a divergence in the underlying methods. As our distance estimates used the same prior choices as \cite{BailerJones2021}, the primary differences between the results arose from our NN providing an improvement over the base Gaia parallaxes. Thus, this bias maps the regime where the our distances were more weakly constrained by the statistical prior distribution than in \cite{BailerJones2021} work. Beyond this region, where parallax measurements became too noisy for our NN-based approach to improve upon, the two methods again converge as the prior distribution comes to dominate the distance estimates.

\subsubsection{AstroNN Distance Comparison}

Finally, we compared our distance estimations to those calculated by \citet{Leung2019b} with the AstroNN machine-learning package. The AstroNN package is based on similar neural-network algorithms to our own, and uses APOGEE DR17 spectral data to estimate astrophysical parameters such as stellar abundances, ages, and distances. Therefore, we used these distances as an independent sample from which we could draw comparisons to our own results.

\begin{figure}
	\centering
	\includegraphics[width=.45\textwidth, trim={0, 0, 40, 40}]{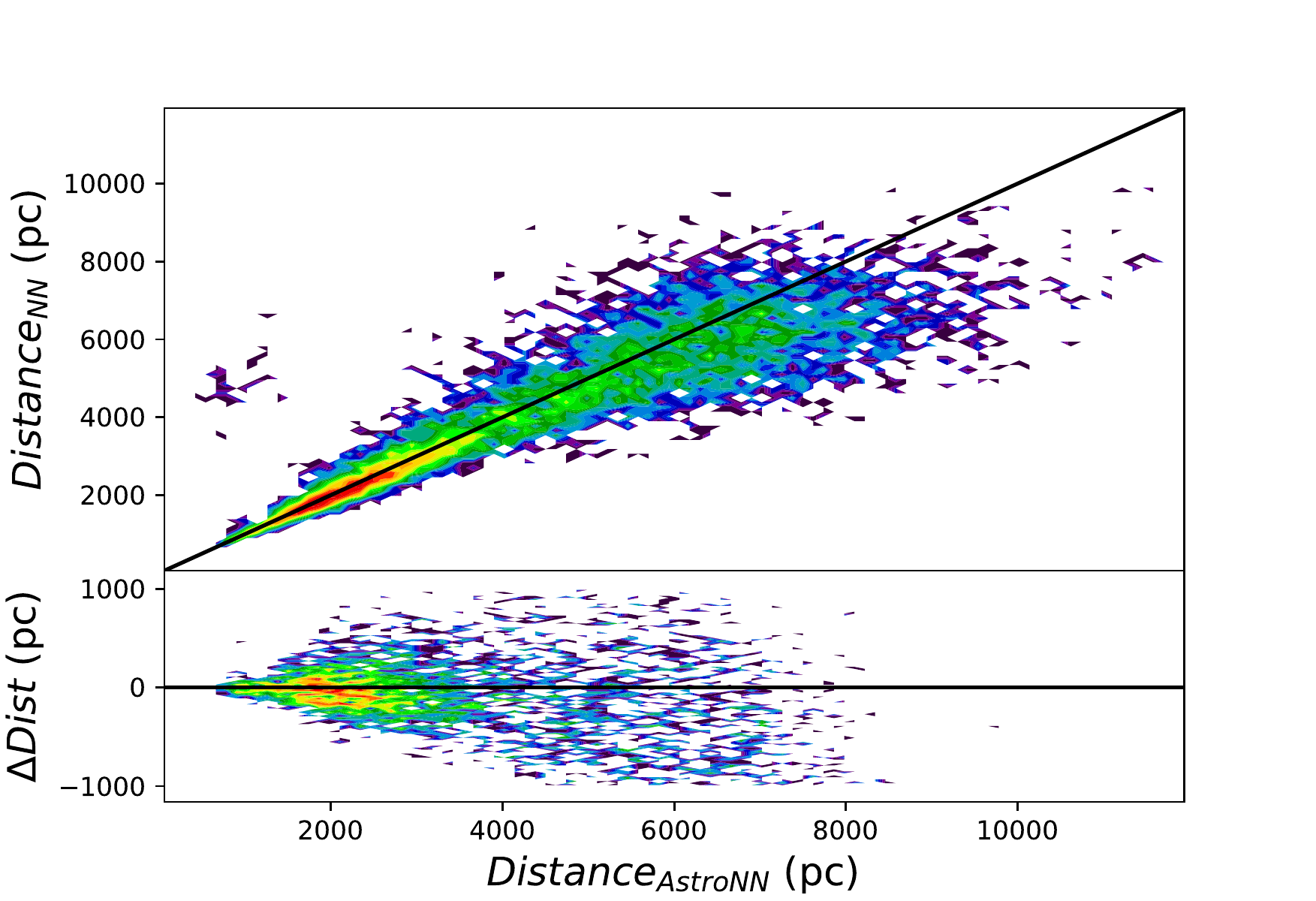}
    \caption{Plot of NN distances from our sample vs. distances from AstroNN (\emph{above}), with comparison residuals (\emph{below}). Note, this contour plot is logarithmic, and is not limited to only stars with Gaia radial velocities.}
    \label{fig:fig_fullNN}
\end{figure}

Using a sample of 11,318 common stars (not limited to only those with Gaia radial velocities), we plotted the comparison in Fig.~\ref{fig:fig_fullNN}. It was clear there was a strong correlation between the two methods with narrow deviations. The differences were also symmetric, suggesting no significant systematic errors in our method that had caused notable biasing. However, as the majority of our sample overlap existed at distances less than $\sim4\,\mathrm{kpc}$, a large proportion of sample objects had informative parallaxes. Therefore, we expected to see this strong agreement when comparing these two approaches. Overall, we concluded that our method has very good agreement with the AstroNN distances, and further confirms the reliability of our distance estimates.


\section{Metallicity Estimation}
\label{Feh_Est}

With an accurate measure of distance determined for each object, we applied our method to predict stellar metallicities. 

\subsection{Data Collection}

We built two samples from which we can estimate metallicities: a training (TG) sample, and a photometric-only (PO) test sample.

The PO sample followed the approach detailed in Section~\ref{Dist_data_coll}, drawing astrometric and photometric data from Gaia EDR3 and the 2MASS and UnWISE surveys. We also included the distance estimations determined in Section~\ref{Distance_Est}, and applied the same filtering to ensure Gaia astrometry includes radial velocities for kinematic analysis. This sample acts as our `output' sample, upon which we will be applying our method for predicting metallicities.

Our TG sample contained the data we will use to train our NN algorithms. This was built from matching objects from our PO sample with iron abundance measurements ([Fe/H]) derived from two spectroscopic surveys, APOGEE-2 (SDSS DR16) \citep{Majewski2017, Ahumada2020} and LAMOST DR6 \citep{Cui2012}. This sample covers the magnitude range of $9\leq G\leq15.6$ in the Gaia G-band. This spectroscopic information can then be used as the dataset our NN is trained to estimate from photometric data. We acknowledge that while broadband photometry will be sensitive to overall stellar metallicity, we use spectroscopic iron abundance as an accurate proxy for this value.

We removed objects with poor spectroscopic data by excluding sources with $\sigma_{\mathrm{T_{eff}}}/T_{\mathrm{eff}}>1$ and $\sigma_{\log g}/\log g>1$. As the range of metallicities in the training data crosses [Fe/H] = 0, using fractional uncertainties causes us to filter valid objects with small absolute metallicities. Thus we do not apply this filter to metallicity. We instead incorporate training data uncertainties as part of the NN's training process (as described in Section~\ref{NNsetup}) which accounts for metallicity uncertainties in the training dataset.

Together, these two spectroscopic surveys provided a large sample of objects, mainly due to the large sky region and depths observed by the LAMOST survey. We were therefore confident our training sample had high-quality metallicities with minimal bias from spatially unbalanced datasets. We note that, thanks to calibration between giant stars in LAMOST and APOGEE datasets, the two spectroscopic surveys shared a good agreement with their metallicity observations \citep{Anguiano2018}. Thus, while small discrepancies may occur, we felt confident using the two surveys concurrently. Additionally, in situations where objects appear in both APOGEE and LAMOST, we preferred the higher-resolution APOGEE data and included only this value in our sample.

\subsection{Methodology}

We built our network with architecture described in Section~\ref{NNsetup}, and selected input features constructed from 16 photometric colours and 8 absolute magnitudes (as described in Appendix~\ref{appendix_NNsetup}). For training, we used the input features alongside our TG sample's spectroscopic metallicities to optimise the network to predict metallicities from photometry. 

A major deviation from the method used in Section~\ref{Distance_Est} was the inclusion of an extra weighting term to the network’s loss function which worked to down-weight objects with [Fe/H] $\approx 0$. This aimed to oppose the significant over-abundance of near-solar metallicities in our TG sample. Without mitigation, the network would learn this imbalance as a trend in the data, and return values which follow this bias. Thus, we would have expected the algorithm preferentially return metallicities close to zero, as (when averaged over the entire dataset) these predictions would be generally accurate. 

Our weighting term took the form of a linear multiplier on the network's loss function. Modifying Eq.~\eqref{eq:one}, with this weighting term as $W = |[\mathrm{Fe/H}]| + C$ (where $C$ is a constant), the weighted loss function is
\begin{equation}
    J(y_{i}, \hat{y}_{i}) = \frac{1}{n} \sum_{i=1}^{n} \frac{W}{2} (y_{i} - \hat{y}_{i})^2 e^{-s_{i}} + \frac{1}{2} s_{i}.
	\label{eq:loss_weight}
\end{equation}

This weighting acted to decrease the `loss' penalty when training on objects with [Fe/H] $\approx 0$, and increased the penalty linearly for objects with much larger or smaller metallicities. Thus, the network put less effort into accurately predicting objects with solar-like metallicities, as penalties were significantly smaller for poor estimations. The weighting was also tuned with the constant, $C$, which changed the minimum (and maximum) weight an object can be allocated. For our training, we selected $C = 0.5$, such that the penalty multiplier for an object with [Fe/H] = 0 was $\times0.5$ and an object with [Fe/H] = -2 was $\times2.5$. We chose this value to increase the network's sensitivity to very low- and high-metallicity objects, while reducing the priority of metallicities between $-0.5 < \mathrm{[Fe/H]} < 0.5$ (the metallicity region of the majority of our TG sample). This ensured that objects with near-solar metallicites still retained a small impact on the network training, while maximising the relative weighting between the high and low ends of the metallicity range.

One small side-effect of this weighting procedure was a reduction in accuracy for objects with [Fe/H] $\approx$ 0, due to the network considering them as lower priority. However, this had a negligible effect on the overall prediction accuracy: the larger population of objects with [Fe/H] $\approx$ 0 somewhat offset this effect, while the improvements to high/low metallicity predictions provided much more significant enhancement.

We also note that the inclusion of the weighting term in equation~(\ref{eq:loss_weight}) may have also caused a small increase in the uncertainties output by the network. As $s_{i}$ incorporates the predictive uncertainty of the NN, the network may have returned slightly larger uncertainties to account for the weighting term. For objects at high- and low-metallicities, which would be most affected by the weighting term, this uncertainty increase would be the most severe. In this case, we would expect the potency of the weighting term's bias-reduction would have been reduced.

The success of this approach was not perfect, as we found that uncertainties still vary with respect to predicted metallicity. As shown in Fig.~\ref{fig:fig_seven}, even with the weighting term included, the prediction uncertainty was far larger for the highest and lowest metallicity objects. However, for the majority of our sample, the uncertainties remained small enough to be sufficient for our purposes.

There are two potential approaches to mitigate this in future work: a more complex weighting criteria, to better reduce the impact of unbalanced data; or observing a greater number of objects with extremely high/low metallicities. While weighting may work to successfully mitigate this issue in some instances, removal of the imbalance altogether would be preferred, which can only be achieved through the latter of these two solutions. 

The use of narrower bands, especially those bluer than in our data, may form a notable improvement over using broad-band photometry alone. The benefits of these bands for measuring stellar parameters has been shown by \citet{SkyMapper_design} and \citet{PIGS}. However, the extreme extinction effects in these bands within regions such as the mid-plane or central bulge adds additional complexities to their inclusion into our dataset.

\subsection{Validation}
\label{sub:validation}
To validate the prediction accuracy, we had two measures: the uncertainty output calculated by the NN, and its performance compared against spectroscopically-determined metallicities.

\begin{figure}
	\includegraphics[width=\columnwidth, trim={0, 0, 40, 0}]{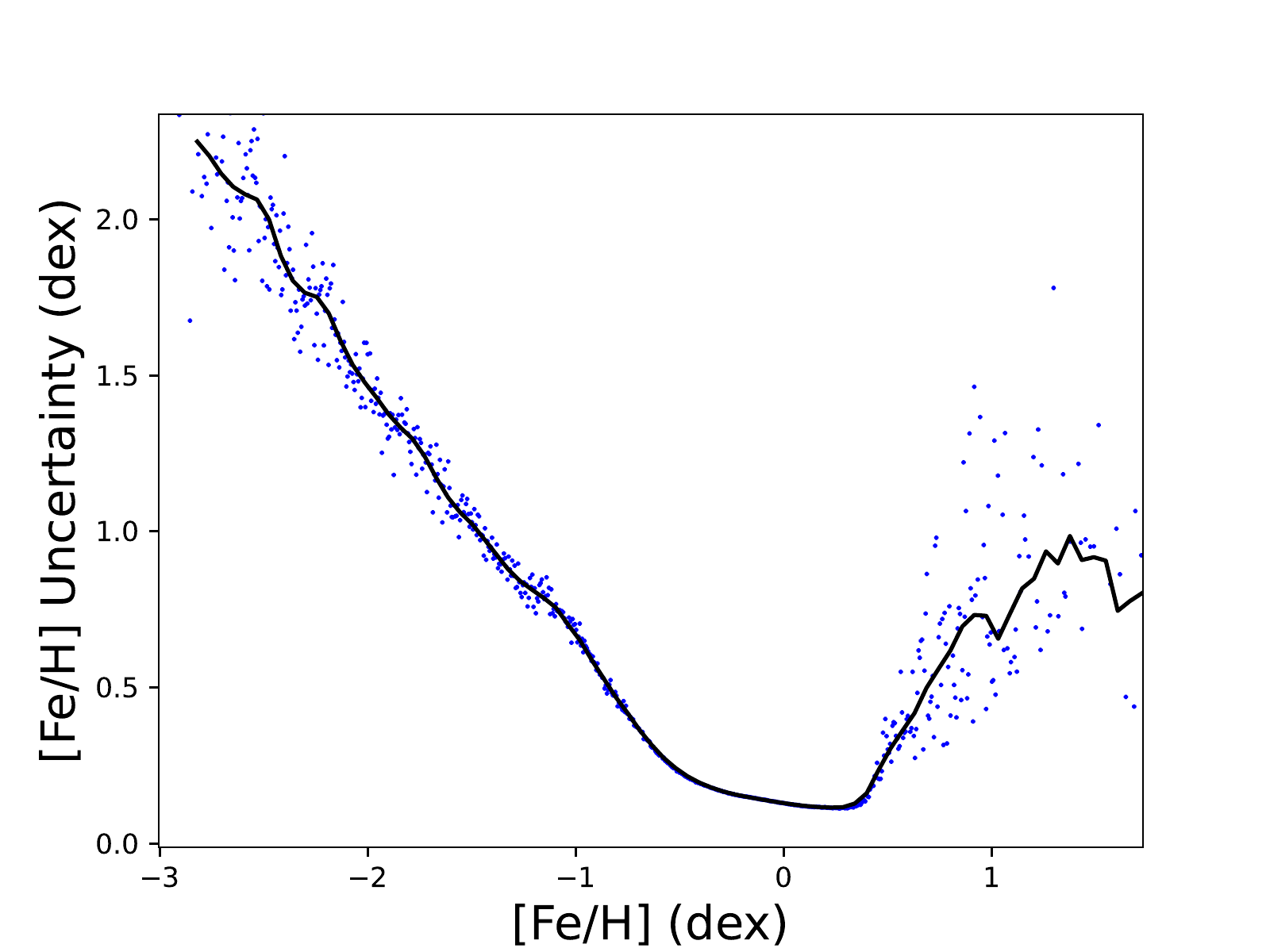}    
	\caption{Plot of metallicity uncertainty against  metallicity for our NN-estimated photometric metallicity values. The lowest uncertainty predictions are those with absolute metallicity close to 0.0 dex (the highest population region), with uncertainties becoming more significant at the edges of our distribution.}
    \label{fig:fig_uncWrtFeh}
\end{figure}

From the network’s uncertainty measure, we found a very high confidence in the metallicity predictions being made. We returned a mean uncertainty output of $\pm0.185\,\mathrm{dex}$ over our entire sample. We show our metallicity uncertainties binned by predicted metallicity in Table~\ref{tab:Feh_unc_tab}. This reiterates the correlation shown in Fig.~\ref{fig:fig_uncWrtFeh}. It is clear that within the range $-0.5 < \mathrm{[Fe/H]} < 0.5$ our predictions perform the best with an uncertainty of $\pm 0.15$, and we have worse performance at low metallicities ($\mathrm{[Fe/H]} < -1.5$). We also found that, while there is a large tail of high uncertainty predictions, these outputs only make up a small fraction of our entire sample: 97.49\% of our PO sample have uncertainties below $\pm0.5\,\mathrm{dex}$. In comparison, the spectroscopic metallicities in the range $-0.5 < \mathrm{[Fe/H]} < 0.5$ have a mean uncertainty of ±0.046$\,\mathrm{dex}$, meaning our best-case metallicities have uncertainties about three times that of the spectroscopic data. We find these uncertainties are comparable to the results of other photometric-metallicity methods, with \citet{Grady2021} finding an uncertainty of ±0.21 ($\mathrm{[Fe/H]} > -0.5$), \citet{Huang2021} finding an uncertainty of approximately ±0.12 ($-0.5 < \mathrm{[Fe/H]} < 0.5$), and \citet{Lin2022} finding an uncertainty of ±0.2 dex. Furthermore, we directly compare our metallicities to those from \citet{Huang2021} and \citet{Lin2022} in Appendix~\ref{appendix_fehcomp}, and find reasonable agreement to these methods.

We note that the range of Fig.~\ref{fig:fig_uncWrtFeh} extends beyond the metallicity range of our TG sample. The minimum metallicity from the spectroscopically measured giant stars was -2.49 dex, and the maximum metallicity being 0.74 dex. Outside of this range, the NN must extrapolate beyond the training data - and thus causes returned uncertainties to be very large. This is most apparent above 0.74 dex, where uncertainties become extremely large beyond the extent of the training data. We therefore recognise that metallicities at the extreme edges of our metallicity distribution should be ignored in further analyses (either by specific cuts to metallicity, or by filtering for extreme metallicity uncertainties).

\begin{figure}
	\includegraphics[width=\columnwidth, trim={0, 0, 30, 40}, clip]{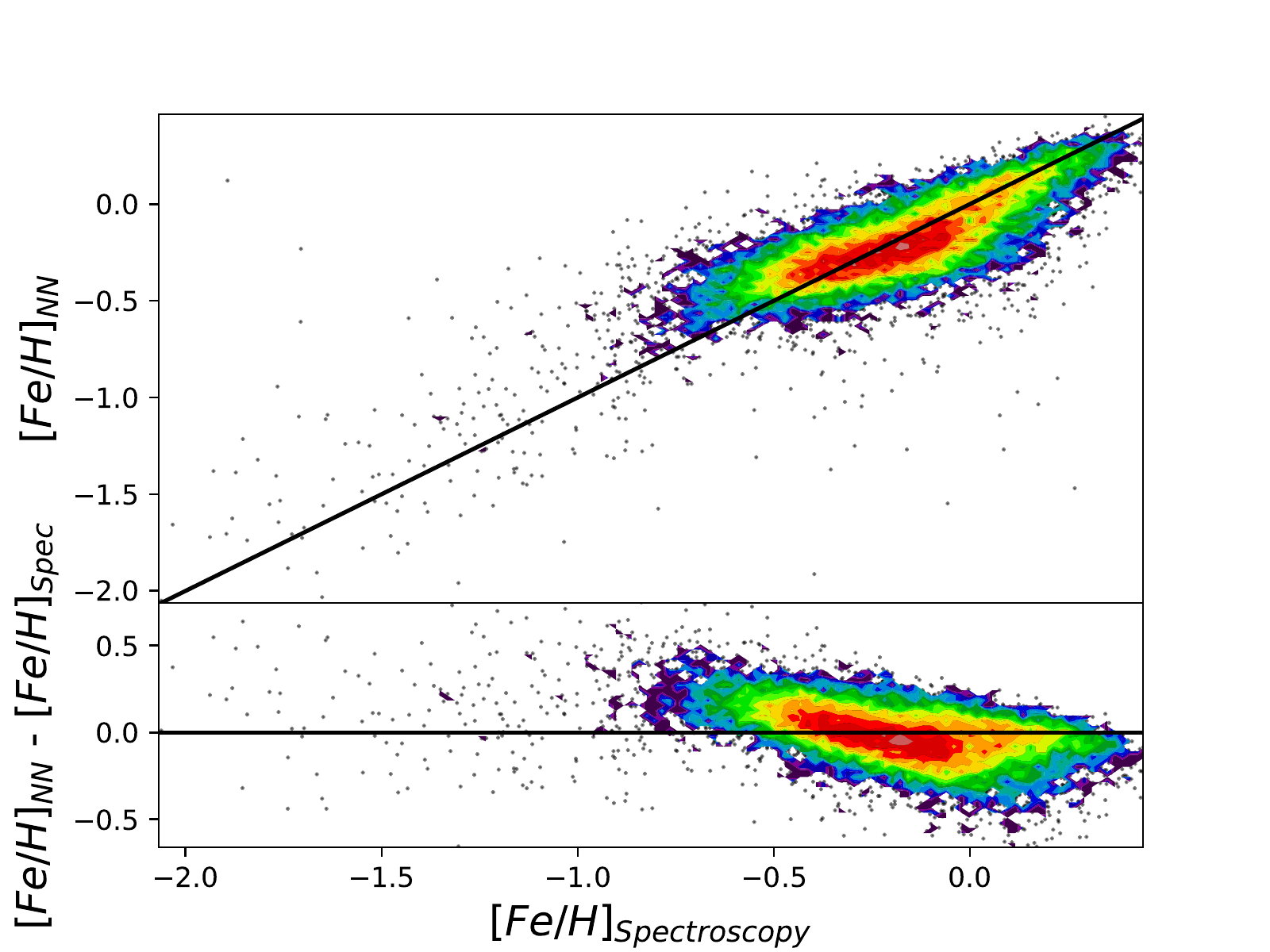}
    \caption{Plot of NN-estimated photometric metallicities from our cross-validation vs. metallicities from spectroscopic observations  (\emph{above}), with comparison residuals (\emph{below}). Note, this contour plot is logarithmic.}
    \label{fig:fig_seven}
\end{figure}

We further compared the predictions made by our NN to metallicities from APOGEE and LAMOST, providing a measure of the `recovery accuracy’ of the network. This worked to cross-check the uncertainty values outputted by the NN, ensuring that the network retains its high accuracy when compared to `true’ data values. This validation is achieved through a method of out-of-bag cross-validation. We selected a fraction of our TG sample to be removed from the network's training process, which we then used to validate the model's predictions. We chose a validation sample split of 15\% of our TG sample, leaving 85\% to train the network. The network's predictions on the validation sample were then be compared to the spectroscopic measurement, with the comparison shown in Fig.~\ref{fig:fig_seven}. We find there is a good correlation between our method and the spectroscopic data, suggesting our approach is successful in accurately reproducing metallicity values. However, we do confirm the minor bias apparent in the residuals at high and low metallicities, with an over-estimation of metallicities below [Fe/H] < -0.5 and a smaller biasing of under-estimated metallicities for high [Fe/H] objects. This `regression to the mean' effect is a common issue for neural-network algorithms using unbalanced datasets, and so suggests our weighting term has not fully removed these effects. Analyses using lower metallicity objects must take this into account.

\begin{figure}
	\centering
	\includegraphics[width=.45\textwidth, trim={0, 0, 40, 0}]{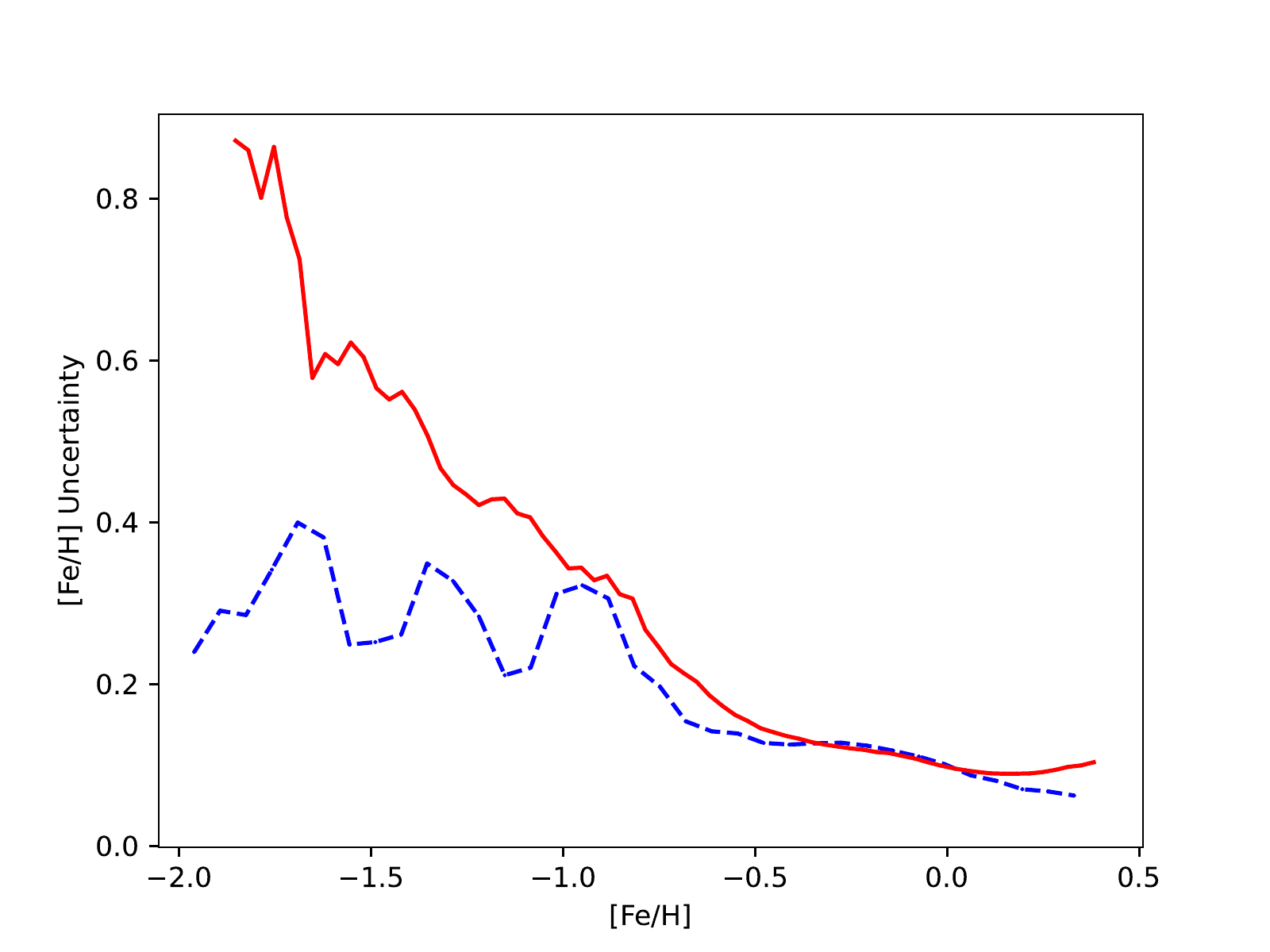}
    \caption{Plot of NN metallicity uncertainties (solid red) and the residuals between the NN and spectroscopic metallicities (blue dashed) from Fig.~\ref{fig:fig_fullNN}'s lower panel, plotted against absolute predicted metallicity. These uncertainties diverge notably from the trend shown in Fig.~\ref{fig:fig_uncWrtFeh}, as these are objects from our spectroscopically-matched TG sample --- and thus, tend to be closer and brighter than many objects in our output sample. }
    \label{fig:UncVsResids}
\end{figure}

We finally analyse the effect the weighting term may be having on the predicted metallicity uncertainties, as noted in the previous section. We compare the NN's output uncertainties to the residual scatter in the lower panel of Fig.~\ref{fig:fig_seven}. If the network is predicting larger uncertainties due to inclusion of the weighting term, we would expect the the output uncertainties to be much larger than the scatter in the residuals. We plot this in Fig.~\ref{fig:UncVsResids}. Note that we have significantly fewer objects at [Fe/H] < -1 (137 objects) than for [Fe/H] > -1 (15,430 objects), and so our trends are poor beyond this threshold. This figure shows clearly that, for the metallicity range where we have large numbers of objects, we see a good agreement between NN uncertainties and residual scatter.  Thus, we conclude that the weighting term does not appear to be causing the NN uncertainties to be output significantly larger than expected. Furthermore, we note that the uncertainties shown in Fig.~\ref{fig:fig_uncWrtFeh} and Table~\ref{tab:Feh_unc_tab} may be over-estimated at the low-[Fe/H] regime, as they are significantly larger than we would expect from the residual scatter trend.
\begin{table}
    \centering
    \caption{Table of mean metallicity uncertainties for binned  metallicity ranges. Note that the top-most row shows the mean metallicity uncertainty over the entire sample.}
    \label{tab:Feh_unc_tab}
    \begin{tabular}{cclc}
        \hline
        {[}Fe/H{]} Bounds                        & {[}Fe/H{]} Unc. & Obj. Counts \\ \hline
        $-3.5 <  \mathrm{[Fe/H]} < 1.5$  & ± 0.185 & 1,697,077        \\ \hline
        $-3.5 < \mathrm{[Fe/H]} < -2.5$ & ± 4.042 & 713        \\ 
        $-2.5 < \mathrm{[Fe/H]} < -1.5$ & ± 1.713 & 12,392        \\ 
        $-1.5 < \mathrm{[Fe/H]} < -0.5$ & ± 0.404 & 145,177        \\ 
        $-0.5 < \mathrm{[Fe/H]} < 0.5$ & ± 0.150 & 1,538,047         \\ 
        $0.5 < \mathrm{[Fe/H]} < 1.5$  & ± 0.537 & 736         \\ 
        $1.5 < \mathrm{[Fe/H]} < 2.5$   & ± 1.009 & 12        \\ 
        \hline
    \end{tabular}
\end{table}

\section{Results and Analysis}
\label{section::results}

With the completion of the metallicity estimation, we returned our PO sample of 1,689,885 objects with: Gaia astrometry; eight photometric colours from Gaia, 2MASS, and WISE; kinematic information from Gaia proper motions and radial velocities; and photometric metallicity estimations. Furthermore, we calculated three-dimensional Galactocentric coordinates and velocities based on our distance estimates. Using a right-handed coordinate system, we converted Gaia astrometry (sky positions and velocities) into Galactic positions and velocities. In this system, the $X$ axis is along the Sun-Galactic Centre (GC) direction with positive towards the GC. The longitudinal axis, $Y$, sits perpendicular to $X$ along the Galactic plane, with positive $Y$ in the direction of positive Galactic longitude. The vertical axis, $Z$, is directed out of the Galactic plane with positive towards Galactic north. All axes have their origin at the GC.

We applied our catalogue to determine the out-of-plane metallicity gradient of the Galactic bulge, and to identify the vertex angle of the Milky Way's bar from stellar kinematics and metallicity.

\subsection{Vertical metallicity gradient in the Galactic Bulge}

The presence of a metallicity gradient, vertically out of the Galactic plane, in the region of the bulge has been identified in many previous studies. This gradient is suggested by some to be the effect of overlapping populations within the bulge region \citep{Barbuy2018}. These intersecting structures include bulge and bar populations, as well as the surrounding disk and halo structures. As we observe away from the Galactic plane, we see the changing influence on each of these independent components, which creates a gradient in the observed metallicity distribution. Alternatively, other work proposes that this gradient instead forms from the kinematic separation of different populations during the formation of the bulge and bar \citep{Debattista2017}. Due to bursts of star formation during bulge formation, populations of metal-poor and -rich stars can become separated kinematically into hotter and colder velocity distributions. This causes a metallicity gradient to be observed, without the need for distinct, overlapping populations. As summarised by \citet{Ness2016}, gradients have been observed in past literature of around $-0.45$ dex/kpc \citep{Ness2013}, with some methods observing as low as $-0.6$  dex/kpc \citep{Zoccali2008} and as high as $-0.35$  dex/kpc \citep{Minniti1995}.

To draw this trend from our data, we first defined our selection region of the `bulge'. Using our 3D Galactocentric Cartesian coordinates, we defined our bulge region to be within 2.5kpc radius (along the plane) of the Galactic centre -- selecting a cylindrical volume centred on the Galactic centre. We also applied filtering on selected objects, removing stars with metallicity uncertainties greater than ±0.5dex, positional uncertainties greater than ±1kpc, and velocity uncertainties greater than ±250km s$^{-1}$. We note that the filter on metallicity uncertainty will ensure we are only selecting objects with `good' metallicity estimations, but will also introduce a bias into the gradient observed. Filtering out objects with metallicity uncertainty greater than ±0.5 dex will predominantly remove objects with [Fe/H] < -1 dex and [Fe/H] > 0.5 dex. Thus, the trends we observe in metallicity will potentially ignore populations of high- or low-metallicity stars that would otherwise shift the mean metallicity at a chosen position in the Galaxy, and cause our recovered gradient to be under- or over-estimated.

\begin{figure*}
	\centering
	\includegraphics[width=.45\textwidth, trim={0, 0, 30, 30}, clip]{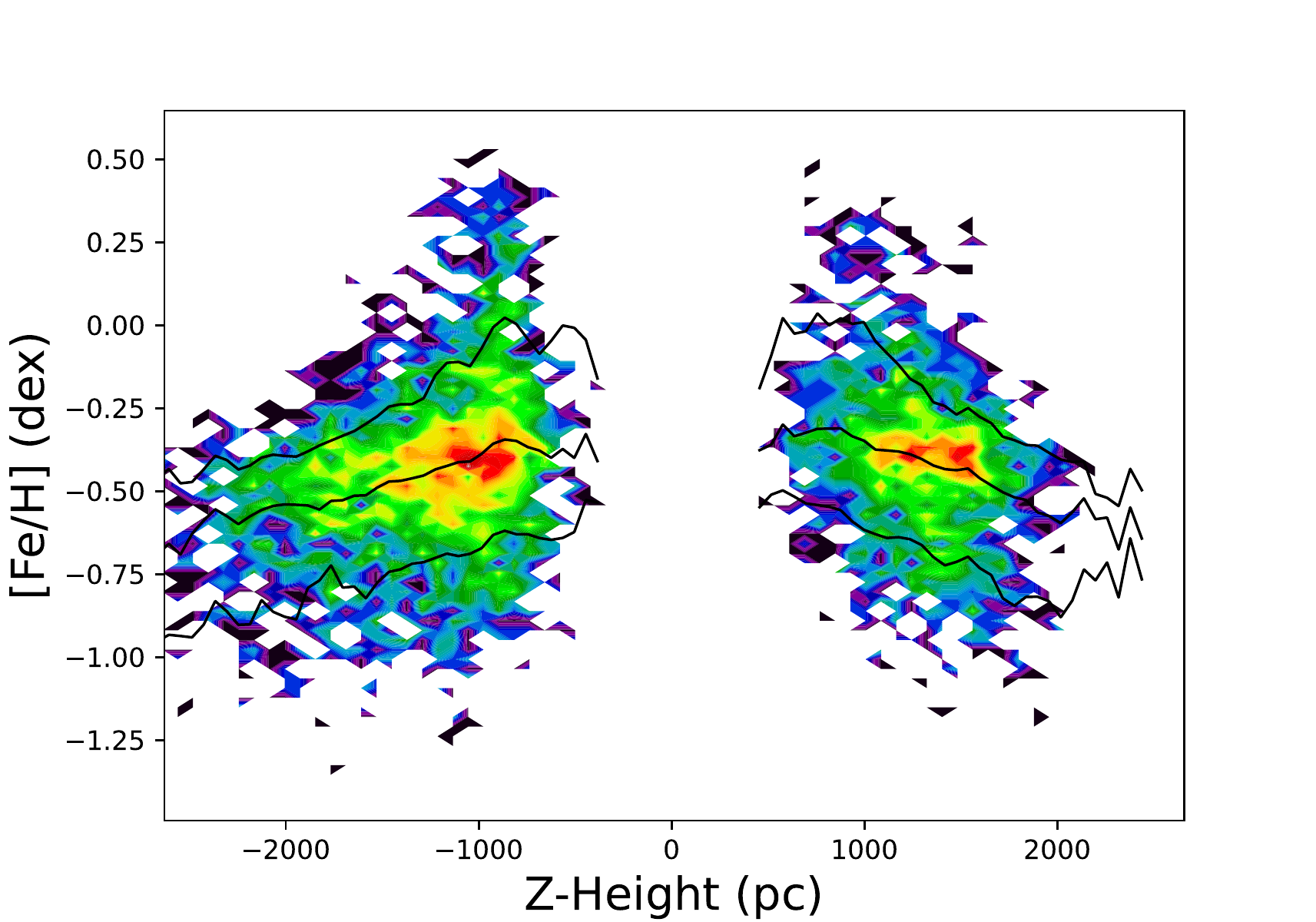}
	\includegraphics[width=.45\textwidth, trim={0, 0, 30, 30}, clip]{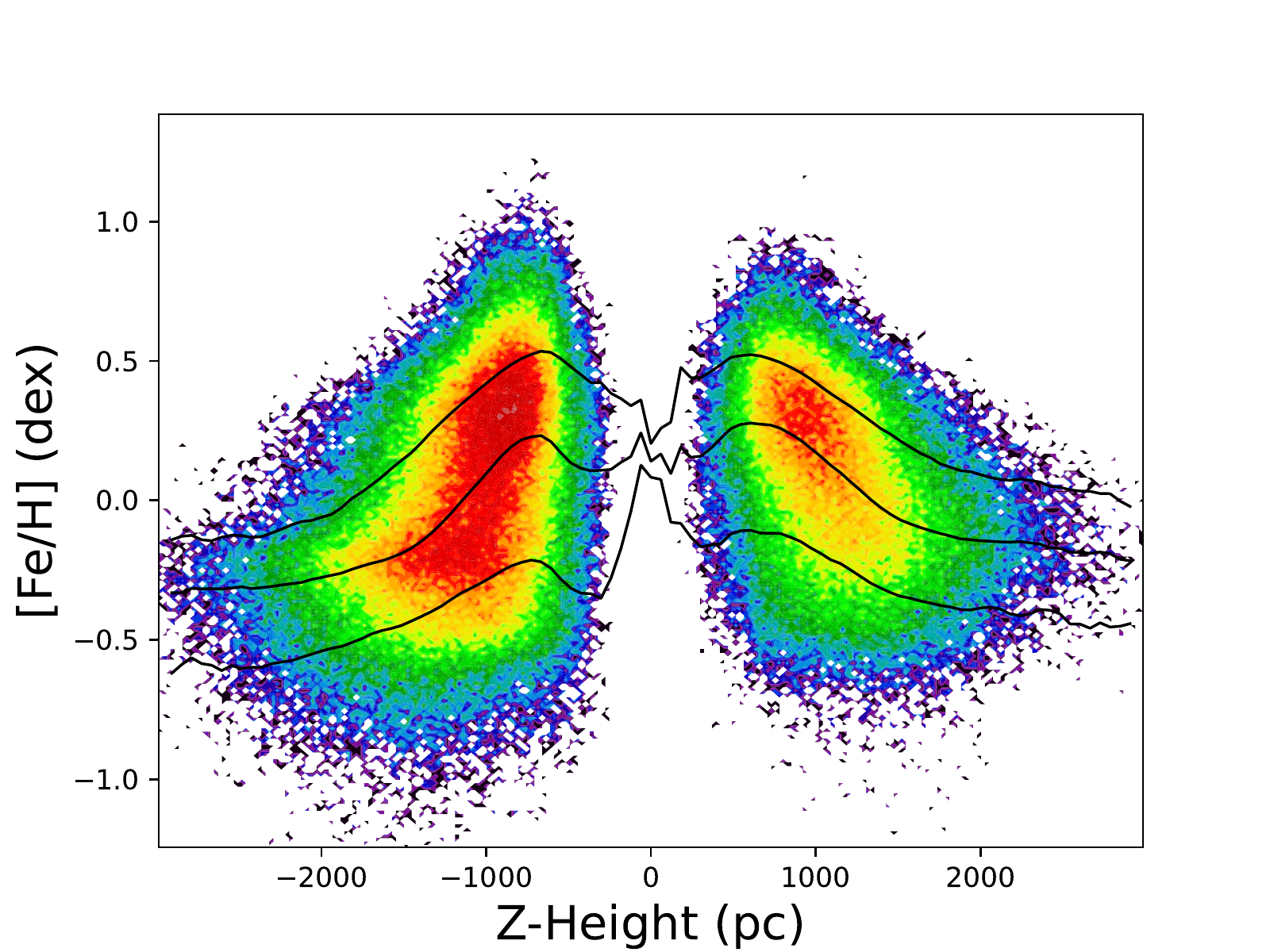}
	\includegraphics[width=.45\textwidth, trim={0, 0, 30, 30}, clip]{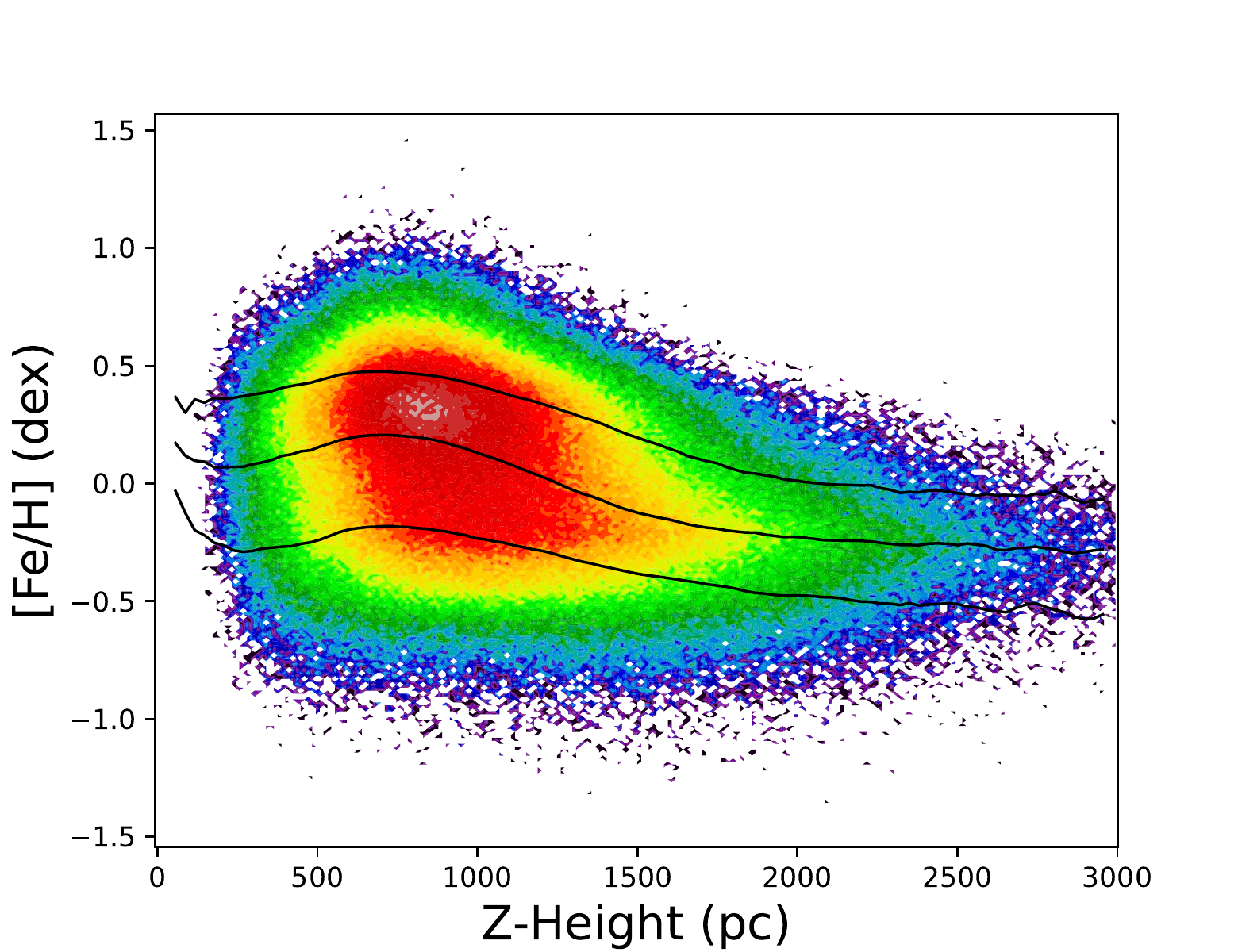}
    \caption{Logarithmic contour plot of median metallicity against an object's height, $Z$, for objects within the bulge selection volume. We plot objects with radial velocitiy information (\emph{left}), objects without radial velocities (\emph{right}), and objects without radial velocities plotted against \emph{absolute} $Z$-height (\emph{bottom}). For the \emph{bottom} panel, we find the median curve to peak at approximately 560pc with a metallicity of 0.173dex. Note that the upper and lower curves are the 84th and 16th percentiles respectively.}
    \label{fig:zfeh_combo}
\end{figure*}

Initially, we plot stellar metallicity against object height above/below the Galactic plane, $Z$, for the PO sample in the \emph{left} panel of Fig.~\ref{fig:zfeh_combo}. It is clear the trends visible are very noisy, with a large uncertainty across the range of $Z$-values shown. We find this is a limitation due to the small number of objects with Gaia radial velocity information within the volume, limiting us to only 22,280 objects. This small sub-sample leads to a large scatter in median metallicity with $Z$-height, and reduces the strength of trends we can draw. Fortunately, we do not need velocity information to draw a positional metallicity gradient, and thus we could remove this requirement when collecting our dataset and expect to see more objects within the sample volume.

Including objects without radial velocities, we applied our NN to estimate metallicities for the larger sample. This is shown in the \emph{middle} panel of Fig.~\ref{fig:zfeh_combo}, which mirrors the \emph{left} panel while showing a much stronger trend across the $Z$-height range. As we expect the gradient to be symmetric above and below the plane, we further plot the median metallicity against the absolute $Z$-height in the \emph{right} panel of Fig.~\ref{fig:zfeh_combo}, which increases the strength of observed trends and further constrains the level of uncertainty in a measured gradient.

We also note the clear gradient inversion visible within 500pc of the plane. This appears to retain the well-constrained uncertainties between approx. 250pc and 500pc, before the trend becomes extremely scattered towards the mid-plane. This is unexpected, as \citet{Gonzalez2016} note that many past works with spectroscopic data have recovered a smooth metallicity-height relation, from lower metallicity objects far from the plane, and higher metallicity objects towards the mid-plane. 

This metallicity gradient change towards lower latitudes has been noted by \citet{Rich2012}, who observed the vertical gradient flattens below a vertical height of 550pc. Furthermore, \citet{Babusiaux2014} found hints that the gradient indeed inverts close to the plane. This is proposed to have been due to early-forming stars becoming trapped in the inner regions of the Galaxy as it formed, and remaining bound in the mid-plane during bar buckling. Alternatively, this low-metallicity core population may be the result of metal-poor gas being funnelled into the bulge by the bar, forming this metallicity inversion towards the mid-plane. We therefore find our results agree with these past findings, and confirm the presence of a low-metallicity population towards the mid-plane.

We compute a metallicity gradient between $700\mathrm{pc} \leq Z \leq 1600\mathrm{pc}$, and return a value of $-0.5278 \pm 0.0022$ dex/kpc (outwards from the galactic plane). It is useful to note that we are assuming a linear relationship between metallicity and z-height within the quoted range only, and so does not account for the gradient flattening at values of $Z$ outside of our selection.

We find a vertical metallicity gradient that is well within the literature range of values, although towards the steeper end. This suggests our observed metallicity distribution diverges significantly form that found by \citet{Minniti1995} and slightly from that of \citet{Ness2013}. We suggest that such a discrepancy is expected between our method and those that use spectroscopic data. Due to the selection criteria used by spectroscopic surveys, we expect to find our photometric-based data to be sampling a slightly different stellar distribution, and return a slightly different metallicity gradient. 

Furthermore, we note that the metallicity bias described in Section~\ref{sub:validation} may have also biased our recovered gradient. As we expect low-metallicity populations to be more common at higher latitudes, we expect the metallicity overestimation bias to have a stronger impact further from the plane. This would cause our gradient to be measured as shallower, as the mean metallicity at higher latitudes would be increased, while the metallicity of the mid-plane would remain mostly unchanged.

Overall, our main conclusion from this analysis remains that our data has successfully returned positions and metallicity estimations, which accurately trace known abundance trends within the Milky Way.

\subsection{The vertex deviation of the bar}
\label{subsub:bar_kine}

As we have radial velocities from Gaia DR2, we have also been able to use our catalogue to analyse the kinematics of the Galactic bar-bulge. One quantity useful for probing the kinematic properties of the bar is the vertex angle or vertex deviation, that is the angle of the major axis of the velocity ellipsoid relative to the Galactic centre direction giving an indication of the orientation of the bar \citep{Zhao1994}.

Vertex angles, $l_v$, are defined as
\begin{equation}
    l_{v} = \frac{1}{2} \arctan\left(\frac{2\sigma_\mathrm{XY}^{2}}{\sigma_\mathrm{X}^{2} - \sigma_\mathrm{Y}^{2}}\right)
	\label{eq:vert_angle}
\end{equation}
where $\sigma_\mathrm{X}^{2}$ is the velocity dispersion on the Galactocentric X axis, $\sigma_\mathrm{Y}^{2}$ is the dispersion in the Y axis, and $\sigma_\mathrm{XY}^{2}$ is the correlation term. The angle is calculated from the Sun-Galactic Centre (GC) line such that the value is within $|l_{v}| \leq 45^{\circ}$. The angle is positive in the direction of positive Galactic longitude (anticlockwise rotated bar), and negative in the direction of negative longitude (clockwise rotated bar). For an axisymmetric velocity distribution, the vertex angle is ill-defined (as the major and minor axes are equal) and we would expect any measurement to be unconstrained. 

Existing literature has measured this value, and found a bar-like signal for metal-rich bulge objects. \citet{Zhao1994} note a vertex angle of $-65^{\circ} \pm 9^{\circ}$ (for [Fe/H] $\geq$ 0.0), while \citet{Babusiaux2010} measures an angle of $-32^{\circ} \pm 9^{\circ}$ (for [Fe/H] $\geq$ 0.3). Both studies also found that low-metallicity objects show a high-scatter, near-zero vertex angle --- and thus not a bar-like signal. This suggests spherical or disk-like rotation in these metal-poor populations.

We note that, for our analysis, we use our Galactocentric coordinate system (X/Y/Z) to determine the angle of the velocity ellipsoid, rather than the usual Galactic coordinates (r/l/b). For the sky region of interest, these are broadly equivalent, but using this definition does alter the returned `vertex angle' in comparison to past literature. We selected these coordinates as it ensures all object velocity vectors have parallel axes, which is not the case when using Galactic coordinates across large sky regions. Note that the orientation of our axes are described at the end of Section~\ref{section:Method}.

For this method, we developed a Bayesian inference process using a STAN implementation in Python \citep[CmdPyStan,][]{standevteam2021}. We constructed a Markov-Chain Monte-Carlo method (MCMC), which accepted three-dimensional velocity vectors, $\boldsymbol{v}_i$, (in the XYZ coordinate system) and corresponding velocity uncertainty covariance matrices, $\Sigma_{\mathrm{unc}}$ and evaluated the log-likelihood as a two-component Gaussian mixture model with distribution means, $\boldsymbol{\mu}_{n}$, and covariance matrices $\Sigma_{\mathrm{v},n}$. Such a mixture model allows us to isolate a minor `anomalous' component from the data, and return a stronger signal of interest. For this method, we evaluated the log-likelihood of each Gaussian component as
\begin{equation}
    \ln\mathcal{L}_{n} = -\frac{1}{2}\sum_i\Big( (\boldsymbol{v}_i-\boldsymbol{\mu}_{n})^\mathrm{T}(\Sigma_{\mathrm{v},n}+\Sigma_{\mathrm{unc},i})^{-1}(\boldsymbol{v}_i-\boldsymbol{\mu}_{n}) + \ln|\Sigma_{\mathrm{v},n}+\Sigma_{\mathrm{unc},i}|\Big).
\end{equation}
The two components are thus evaluated to assign member stars, with the two component distribution given by
\begin{equation}
    \mathcal{L}_\mathrm{total} = \lambda \mathcal{L}_{a} + (1 - \lambda) \mathcal{L}_{b},
\end{equation}
where $a$ and $b$ denote the two components, and $\lambda$ is a ratio of the two components where $0 \leq \lambda \leq 1$, and $\mathcal{L}_\mathrm{total}$ is the overall log-likelihood.

From the major component in the mixture model, we infer the mean velocity, $\boldsymbol{\mu}$ and the velocity ellipsoid, $\Sigma_\mathrm{v}$, which has components
\begin{equation}
    \Sigma_\mathrm{v} = \begin{pmatrix}
    \sigma_\mathrm{X}^2&\sigma_\mathrm{XY}^2&\sigma_\mathrm{XZ}^2\\
    \sigma_\mathrm{XY}^2&\sigma_\mathrm{Y}^2&\sigma_\mathrm{YZ}^2\\
    \sigma_\mathrm{XZ}^2&\sigma_\mathrm{YZ}^2&\sigma_\mathrm{Z}^2\end{pmatrix},
\end{equation}
from which the vertex angle can be calculated using equation~\eqref{eq:vert_angle}. As the MCMC method returns a distribution of covariance matrices, we output a distribution of vertex angles. From this, we calculated a median value and percentile uncertainties for our vertex angle.

We selected our sample as an on-sky region, with $|l| \leq 5^{\circ}$ and $|b| \leq 10^{\circ}$, and limited to distances between 6 kpc and 10 kpc. This forms a volume approximately 700pc wide on the Y axis, 1.5kpc on the Z axes, and 2kpc deep in the X axis. We note here that this volume-based sample selection is strongly affected by the distances, and thus the distance uncertainties, reported for each star. We therefore filtered our objects for only those with distance uncertainties smaller than $\pm$1kpc, to limit the effect of non-bulge/bar objects with large uncertainties being included in the selection. We also applied filters on extreme metallicity uncertainty ($[\mathrm{Fe/H}]_\mathrm{unc}$ < 1.0), extreme Galactocentric velocity uncertainty (velocity unc. $<250\,\mathrm{km\,s}^{-1}$), and perpendicular axis (Y \& Z) positional uncertainty (position unc. < 1kpc). This selection region is strongly limited by the maximum depth of objects with radial velocity information, which causes our sample to be predominantly objects on the near-side of the bulge, with far fewer objects at greater distances.

\begin{figure}
	\includegraphics[width=\columnwidth, trim={0, 0, 30, 30}, clip]{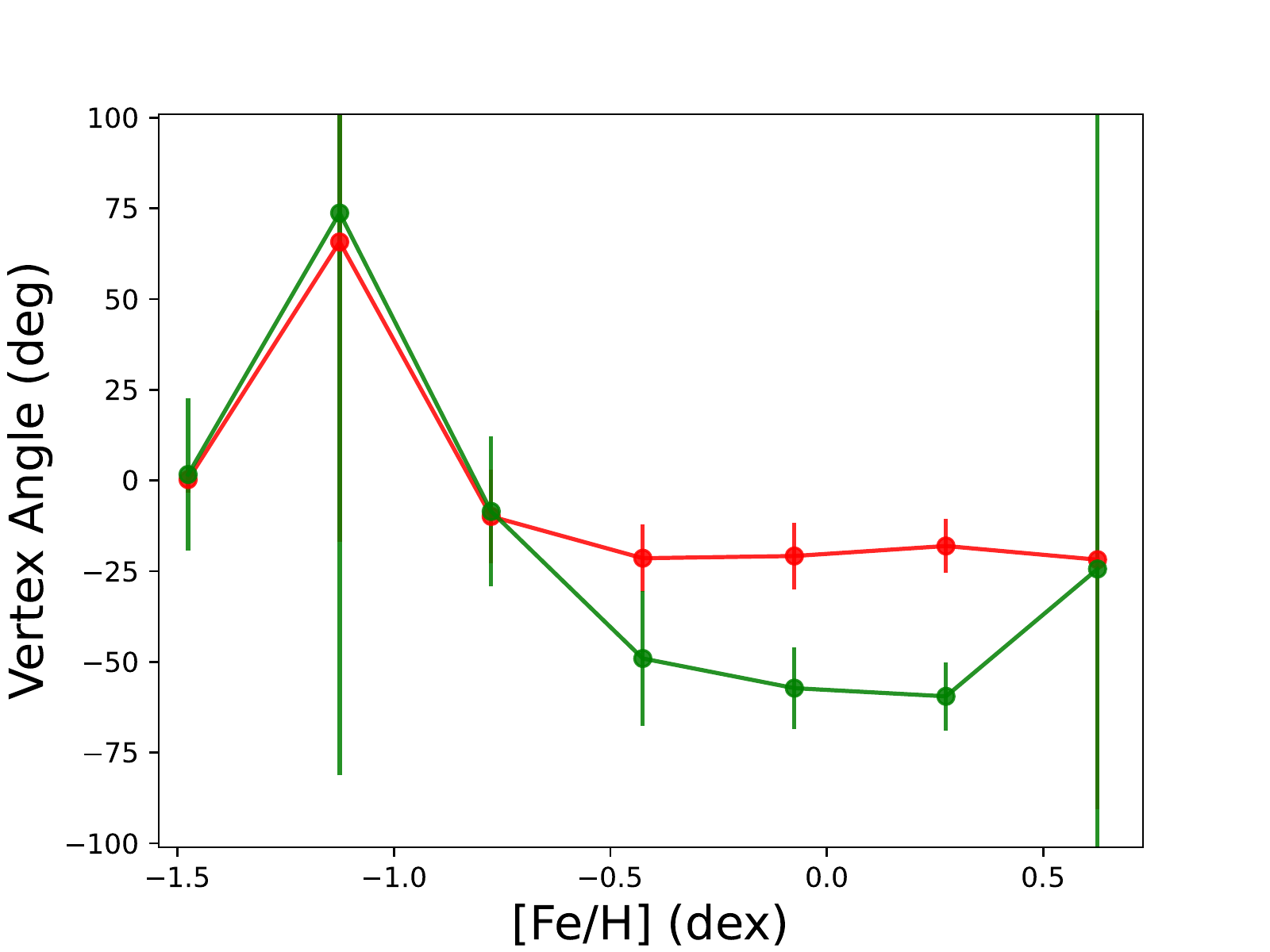}
    \caption{Plot of vertex angle calculated for seven metallicity bins, drawn from $\sim$4,200 objects within the selected bulge region. The angles plotted in \emph{red} are measured with the velocity ellipsoid centred at the mean velocity of the data, while angles plotted in \emph{green} are measured with the velocity ellipsoid centred at the velocity of the Galactic centre. The bin objects counts, from low- to high-metallicity, are: 4, 73, 600, 1882, 981, 418, \& 56.}
    \label{fig:feh_bar_angle}
\end{figure}

To identify the cutoff between high- and low-metallicity samples, we split our sample by metallicity into bins of width 0.35 dex between the range of -1.65 dex and +0.8 dex. Due to the associated uncertainties at extremely high- and low-metallicities, we observe very few objects with $[\mathrm{Fe/H}]_\mathrm{unc}$ < 1.0 outside of this range. We note that for objects beyond the range -0.5 < [Fe/H] < 0.5, the metallicity uncertainty is larger than the 0.35 dex bins we use in Fig.~\ref{fig:feh_bar_angle}. We therefore expect the bins within this `good' range of metallicities to be accurately binned with predominantly objects within the bin range and little contamination. However, for metallicity ranges with higher mean metallicity uncertainties, we expect contamination to be higher between bins. This would cause us to return angles with large uncertainties in these bins, as the contaminant objects will bring a larger distribution of object kinematics.

The vertex angle was calculated within each bin. This binned calculation is plotted in Fig.~\ref{fig:feh_bar_angle}. For low metallicity bins, the vertex deviation is approximately zero, whilst for higher metallicities a large negative angle is found. 

We further compare these values against a slightly modified vertex angle calculation, where instead of centering the velocity ellipsoid on the fitted mean velocity of the data, we assume the mean velocity is zero. As our sample appears to be biased towards objects on the near-side of the bulge, we find that high-metallicity objects have a mean velocity dominated by Galactic rotation. Centering the ellipsoid on this mean removes this net motion from our vertex deviation calculations, and gives a better fit to the kinematic data.

However, this is not the only approach to determine the vertex angle from objects kinematics. If our data was more evenly distributed across the bulge region, rather than predominantly on the near-side, we would expect to find the mean of the velocity distribution close to zero in all axes (rather than dominated by galactic rotation). We therefore estimate the vertex angle with the ellipsoid means `zeroed', to emulate the angle we would return from an unbiased sample. This `zeroed' approach will likely be a poorer fit to our biased dataset, but we find the comparison useful to understand the angle we expect to observe with a kinematically unbiased sample.

We note that our initial method with the means estimated by the algorithm is noted as the `fitted' ellipsoid, while the method with ellipsoid means constrained at zero is the `zeroed' ellipsoid.
\begin{figure*}
	\centering
	\includegraphics[width=.48\textwidth, trim={0, 0, 30, 30}, clip]{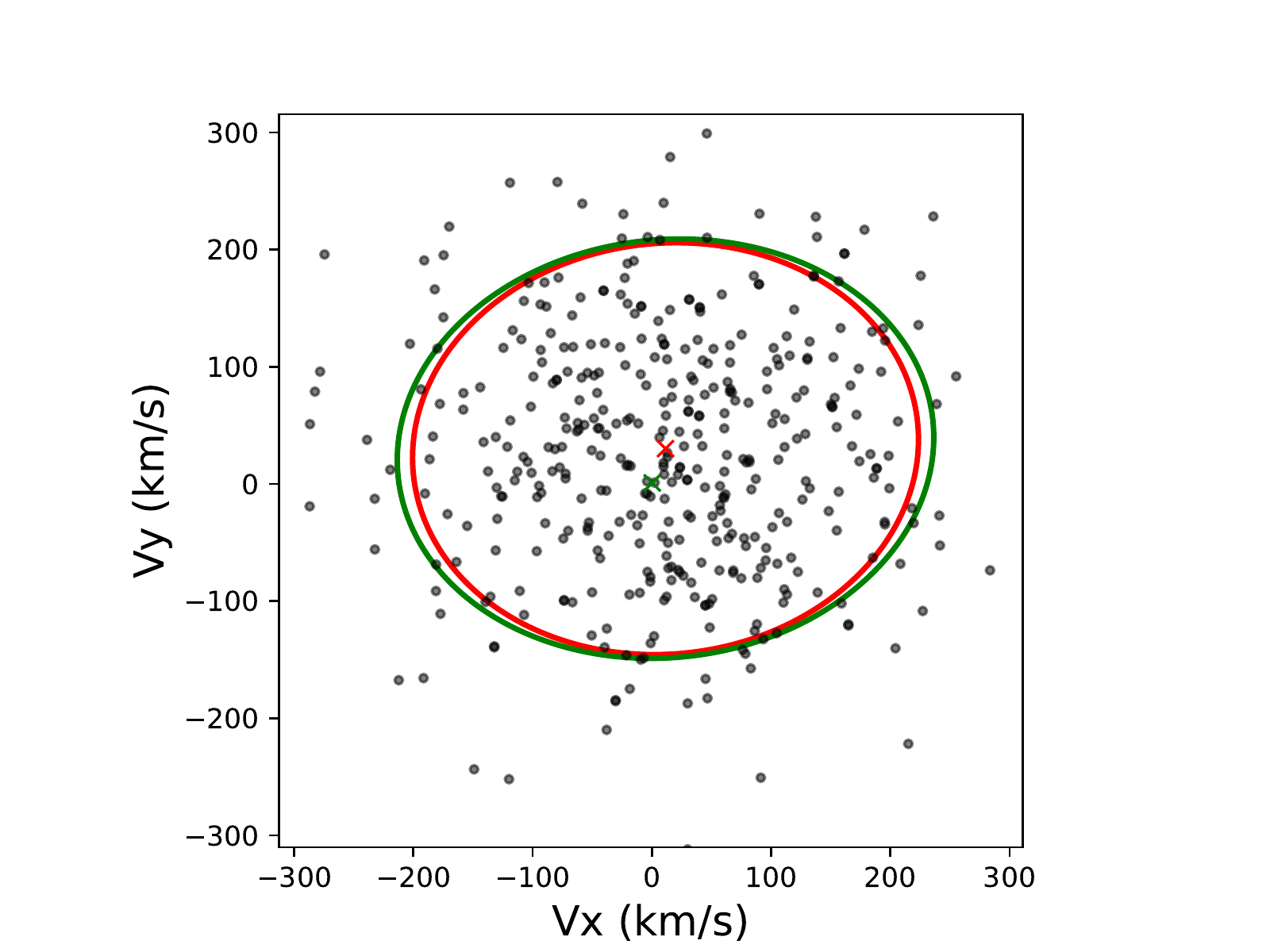}
	\includegraphics[width=.48\textwidth, trim={0, 0, 30, 30}, clip]{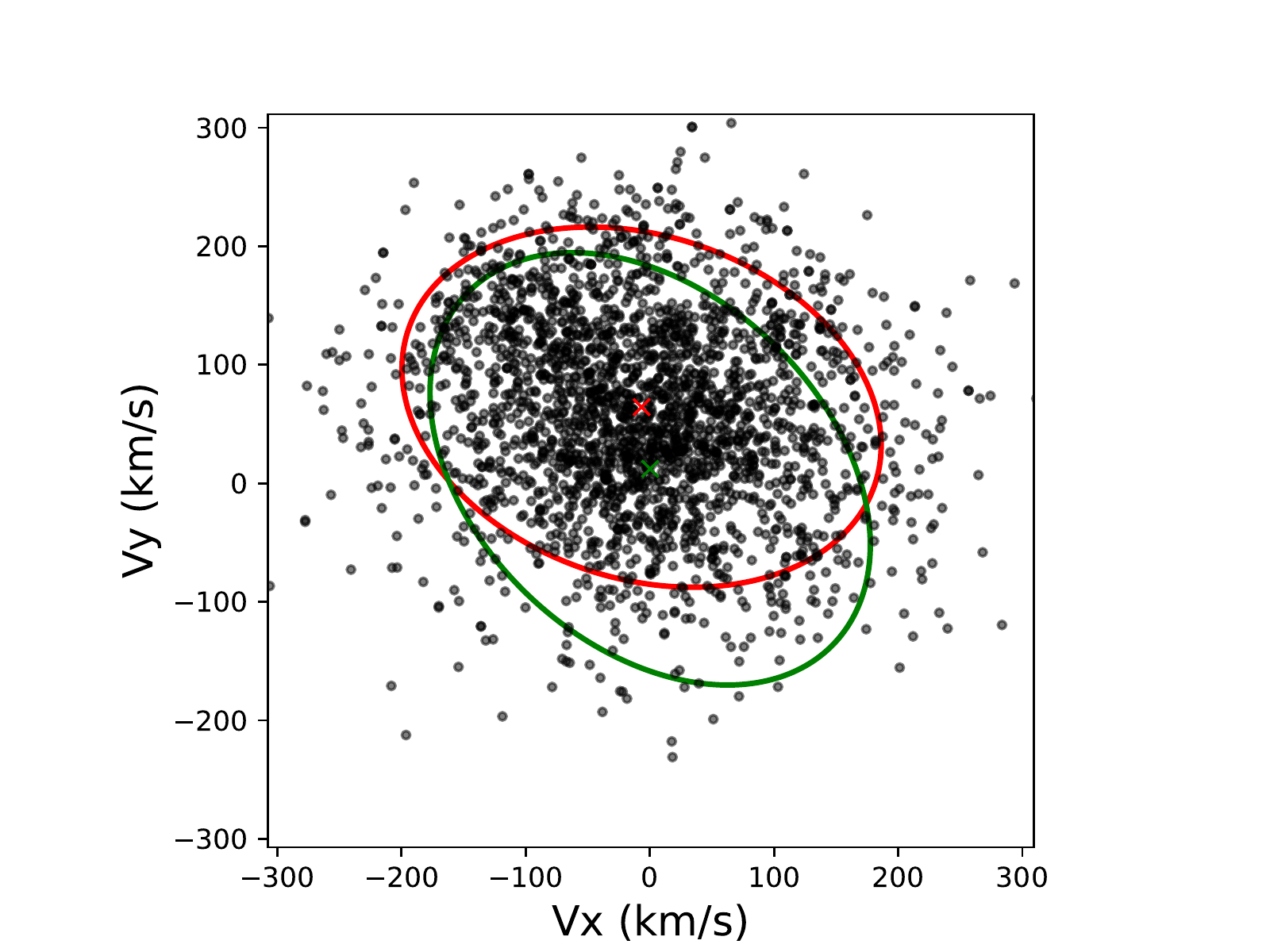}
    \caption{Plot of the velocity distribution on the x (Vx) and y (Vy) axes. The fitted velocity ellipsoids are shown, where the ellipsoids means are fitted to the sample mean velocity (\emph{red}) and the origin of the plot (\emph{green}).}
    \label{fig:vxvy_ellipse}
\end{figure*}

From this binned selection we selected two main samples: a high metallicity bin, and a low metallicity bin. This maximised the number of objects used to calculate the vertex angle, and limited the potential uncertainties from small sample sizes. We selected our low-metallicity bin where $\mathrm{[Fe/H]} \leq -0.7$ and our high-metallicity bin where $\mathrm{[Fe/H]} \geq -0.4$. Using these we retained large samples, and recovered vertex angles with minimised scatter. 

In the region between these two bins, where $-0.7 \leq \mathrm{[Fe/H]} \leq -0.4$, we found a mean angle of $-18.449\pm8.644\,\mathrm{deg}$, which fell between that of the high- and low-metallicity bins and retained a large uncertainty value. This suggested we were seeing an overlap of the two regimes, where scatter in metallicity predictions make it difficult to differentiate the distinct kinematic profiles. We therefore excluded this region from our analysis, and focussed on the selected high- and low-metallicity samples.

We note that contamination between these two bins will be less severe than for the smaller bins used in Fig.~\ref{fig:feh_bar_angle}. As we separate our two bins with the intermediate $-0.7 \leq \mathrm{[Fe/H]} \leq -0.4$ region, there will be few objects with uncertainties extreme enough to contaminate the other bin. The most significant issues will arise for objects with [Fe/H] < -1.5 dex, as their uncertainties become large enough to potentially contaminate the binning. However, as noted previously, we have a very small sample of objects with these very low metallicities. Thus while there may be contamination, we expect this to have a minor influence on the angles calculated.

With the metallicity ranges set, we re-applied the Bayesian model to draw a vertex angle for each of these two samples. These results are shown in Table~\ref{tab:vertex_ang} and show that there was a clear difference between the velocity distribution of the low- and high-metallicity samples. The low-metallicity objects appear to have a small vertex angle with a high uncertainty, suggesting minimal bar-like signal in the data. Conversely, we show a much more negative, low uncertainty vertex angle present in the high-metallicity sample, with an angle of  $-21.29\pm2.74\,\mathrm{deg}$. 

Our results therefore confirm the kinematic split of bulge populations by metallicity. We observe that lower-metallicity bulge objects show more axisymmetric kinematics around the Galactic centre, suggesting they are populations found in the spheroidal-shaped bulge or thick disk. On the other hand, higher metallicity objects show a large, low-uncertainty vertex angle, suggesting these objects instead have a bar-like kinematic structure, and so will be members of the Milky Way's bar population. 

However, the high-metallicity vertex angles are much lower than those found by past works, with our measured angle being around $\sim10$ deg smaller although still within the uncertainties of the measurement from \cite{Babusiaux2014}. This suggests that we either measure a bar that is rotated to a smaller angle than prior works, or a weaker bar-like signal from a more axisymmetric velocity distribution.

We apply our `zeroed' approach to these high- and low-metallicity samples, with both the `zeroed' and fitted mean distributions shown in Fig.~\ref{fig:vxvy_ellipse}. It is clear that, for the low-metallicity sample, both the fitted means and the `zeroed' means trace a similar spherically symmetric distribution, centred on the origin. On the other hand, for the high-metallicity sample, the two distributions diverge significantly, with a vertex angle of $-47.32\pm3.93$~deg, compared to the fitted ellipsoid's vertex angle of $-21.29\pm2.73$~deg. While the fitted ellipsoid is the better fit for the dataset, it is noticeably more spheroidal than the much more extended `zeroed' ellipsoid.

While the `zeroed' approach is a slightly poorer fit to our dataset, this larger angle is much closer to the vertex angle calculated by past works (from -32 deg to -65 deg). This suggests that while our ellipsoid with fitted means is a better fit to our dataset, the velocity bias present means we return an angle that is smaller than expected. Our `zeroed' ellipsoid being closer to the expected vertex angle suggests that mitigation of this bias is necessary to fully recover the bar vertex angle. This can be done either through centering the ellipsoids on zero in all axes, or by building a dataset with greater depth to ensure a more balanced distribution of objects across the bulge. In this case, we would expect the means of the velocity distribution to tend towards zero, and so we would see a distribution closer to that of the `zeroed' ellipsoid.

We do however conclude that we detect a clear difference in vertex angles measured for our high- and low-metallicity samples. The observation that galaxy bar-populations are metal-rich (in comparison to other bulge components) has been discussed by \citet{Wegg2019}, who suggest this describes a formation process where the bar is formed from higher-metallicity, kinematically cool stars which orbit outside of the central bulge, and thus form this separate population within the galactic centre. They also note that a metal-rich bar has also been identified in other nearby galaxies \citep{Gadotti2019}, notably including M31 \citep{Saglia2018}.

Overall, we can confirm the success of our method in recovering this known bar-like signal from objects from photometrically-estimated distances and metallicities. We are also able to highlight the utility of our approach to be applied to structures like the Milky Way's bar, where debates on the metallicity and kinematic distributions are ongoing.

\begin{table}
    \centering
    \caption{Results of vertex deviation calculations of the Milky Way's bulge. The vertex angle is the value returned by the analysis, the Uncertainty is the difference between the 16th and 84th percentiles of the angle distribution. We also include the range (1st to 99th percentile) of the distribution, to illustrate the edges --- and thus the broadness --- of the two prediction distributions.}
    \label{tab:vertex_ang}
    \begin{tabular}{cccc}
        \hline
        [Fe/H] Range (dex)      & [Fe/H] $\leq$ -0.6 & [Fe/H] $\geq$ -0.4 \\
        \hline
        Vertex Angle (deg)      & 6.8526                     & -21.2896                   \\
        Angle Uncertainty (deg) & $\pm$ 16.4271                & $\pm$ 2.7367                 \\
        Angle Range (deg)       & $\pm$ 108.9450                  & $\pm$ 13.2134    
                \\
        \hline
    \end{tabular}
\end{table}

\subsection{Limitations}

We note there is a limitation in our approach to selecting a bulge sample for our two analyses. Our approach in both cases was to select target volumes using cuts in either on-sky Galactic coordinates or Galactocentric positions. These approaches predominantly selected bulge objects within the chosen region of the bulge, and so forms our stellar population of interest. However, we did not make any attempt to isolate any specific population or Galactic component. We therefore note that these selections contain non-bulge populations which overlap the chosen spatial region, such as from the Milky Way's disk or halo. In future work, we hope to include a more robust selection approach, which would account for additional parameters like stellar types or kinematics and allow us to target specific populations with specific analyses. 

Our distance cuts also must account for the potential bias between metallicity and distance. This bias occurs due to low-metallicity stars being brighter than higher-metallicity stars of the same effective temperature \citep{Ahumada2020, Chiti2021}. Low-metallicity stars are then over-selected at greater distances, especially beyond 5 kpc from the Sun (as noted by \citealt{Chiti2021}). However, as our volume-based sample selections collect only a small range of possible distances, the near- and far-sides of out samples will have had approximately similar numbers of over-selected low-metallicity stars. We therefore expect this bias to have only had minor effects on our analyses.

Furthermore, we also note a limitation in how we filtered our samples by metallicity uncertainty. As we wished to focus on how metallicity correlates with object positions and kinematics, we attempted to focus only on objects with `good' metallicity measurements. However, as was noted in Section~\ref{sub:validation}, our metallicity uncertainties vary with absolute metallicity. Therefore any filtering by metallicity uncertainty introduces a bias in our sample, due to removing very high- or low-metallicity objects. This bias was unlikely to cause a major deviation in the trends we observed, as 97.5\% of our output sample has metallicity uncertainties smaller than ±0.5 dex. However, we note that the trends we observe are most strongly applicable to objects with solar-like metallicity, and may not fully account for very high- or low-metallicity populations.

\section{Conclusion \& Discussion}\label{section::conclusions}

Our method to determine metallicity information from photometric information was built on a three-step process:  we first built a neural network algorithm which enhanced Gaia parallax values with photometric information. This allowed us to determine distance estimations with greater accuracy, which we could then bring forwards to predicting metallicities. With accurate distances, we were then able to train our NN model to predict stellar metallicities from APOGEE and LAMOST spectra, allowing the NN to estimate metallicity from photometric colours and absolute magnitudes alone. From this, we could build a sample of objects with Gaia astrometry and metallicity information, allowing for analysis of the positional, kinematic, and metallicity trends in Milky Way populations. Finally, to test our method, we compared against known trends in the Milky Way. Firstly, we measured a vertical metallicity gradient within the Galactic bulge from our data, and compared this to known values in the literature. Then, we used a statistical model to estimate the vertex deviation of different metallicity populations in the Galactic bar.

\subsection{Method Improvements}

Despite our confidence in our results, we still acknowledge there are some outstanding limitations in our method. The primary of these is the amount of data we had available to train the network. Due to the limited depth of Gaia's (DR2) radial velocity data, there was a significant decrease in the number of stars available at large distances. This manifested quite clearly with our analysis in Section~\ref{subsub:bar_kine}, where the limited object numbers in our bulge sample led to a lack of objects at the edges of our metallicity range. Therefore, analysis of the bar's vertex angle required us to choose large bins in our high- and low-metallicity regimes to maintain higher object counts. However, this approach increased the risk of contamination from overlap of the two regimes, as we attempted to maximise the available sample sizes.

Furthermore, our method can be applied to analyses that do not require kinematic information (such as gradients or trends in stellar positions), lacking kinematic information severely reduces the trends and phenomena we can target in future. For example, without velocity information, our methodology would be unable to investigate the properties of bound groupings of stars (i.e. Milky Way structures, accreted sub-structures) where metallicity \emph{and} kinematics are essential to identification and analysis.

With the Gaia DR3 release, the magnitude limit of the radial velocity is fainter \citep{Gaia2021}, and so we expect the number of objects we can return with full kinematic data will increase drastically. This would allow our primary output sample to be significantly larger, and permit us to expand the range of populations we can determine metallicities for. Furthermore, the addition of BP/RP spectra in the DR3 release will provide additional data from which our method can estimate metallicities.

We find another limitation in the lack of a good comparison sample for our metallicity estimations. While we are confident in our abundance estimates thanks to comparisons with APOGEE and LAMOST validation sets, the most robust comparison would be to use an independent survey sample. As we use these two spectroscopic samples as part of our training process, we cannot discern whether our metallicity outputs incorporate the biases or errors from these spectroscopic surveys. Thus, a comparison with an independent survey would ensure these biases could be accounted for. 

In future work, we would be able to build a comparison sample from a selection of current and future surveys. In the immediate future, we could utilise the cross-match between our sample and the GALAH \citep{Buder2019}, RAVE \citep{Steinmetz2020}, or SEGUE \citep{Yanny2009} surveys - each of which is on the scale of 10$^{4}$ to 10$^{5}$ objects, and would have notable overlap with our Gaia-based data. Furthermore, within the next couple of years, the large-scale WEAVE \citep{Dalton2012} and 4-MOST \citep{Jong21019} survey releases would allow us to further compare our method against a wide selection of objects with high-resolution spectroscopic data. The Gaia DR3 release will also have the capability to estimate metallicities directly from BP/RP spectra, which could provide us with an additional sample of metallicities to compare our results against.

Additionally, we further hope to resolve the metallicity imbalance we see in our spectroscopic training data. As the majority of our training sample has near-solar metallicities, this creates a bias in our NN's predictions. While we add heavy weighting to our method to mitigate this, we still find some biasing in our estimations. To more robustly resolve this issue, we can instead augment our existing data with artificial objects, creating a more balanced dataset from unbalanced samples. On one hand, this would require use of generative algorithms, such as variational autoencoders \citep{Kingma2019} or synthetic over-sampling methods \citep{Chawla2011}, which would allow us to generate additional data which would be similar to our input samples. These algorithms could then be used to build samples with significantly more objects with extremely high or low metallicities, and thus work to mitigate the near-solar bias we see currently. 

\subsection{Final Thoughts}

Overall, we can conclude that our NN-based methodology has successfully estimated stellar properties that had previously been difficult to determine without spectroscopic data. Our approach retains high accuracy, with mean uncertainties (for $-0.5 < \mathrm{[Fe/H]} < 0.5$) of $\pm0.15$ dex. We return a catalogue (as described in Section~\ref{section::results}) of 1.7 million Gaia objects with NN-enhanced distances, three-dimensional kinematic information, and accurate metallicity information.

Future works will be able to leverage these results to draw conclusions which require very large samples with stellar abundance information. For example, the identification of substructure within the Milky Way would be an ideal target for our approach, as we have accurate distance, velocity, and metallicity measurements for our large sample. The detection of substructures such the `Gaia-Enceladus Sausage' \citep{Belokurov2018} and `Sequoia' \citep{Myeong2019} merger remnants require the detection of a large number of bound low-metallicity objects to accurately define the structure's origin. While our method does not have the accuracy of spectroscopic approaches, the greater depth and objects counts we return can allow us to find deeper insights into these (and similar) substructures. 

\section*{Acknowledgements}
We would like to thank the anonymous referee for their critiques, and their concise and pragmatic guidance.

J.L.S. acknowledges support from the Royal Society (URF\textbackslash R1\textbackslash191555).
This paper made used of the Whole Sky Database (wsdb) created by Sergey Koposov and maintained at the Institute of Astronomy, Cambridge by Sergey Koposov, Vasily Belokurov and Wyn Evans with financial support from the Science \& Technology Facilities Council (STFC) and the European Research Council (ERC).
This research was supported in part at KITP by the Heising-Simons Foundation and the National Science Foundation under Grant No. NSF PHY-1748958.
This project was developed in part at the 2019 Santa Barbara Gaia Sprint, hosted by the Kavli Institute for Theoretical Physics at the University of California, Santa Barbara.
This work has made use of data from the European Space Agency (ESA) mission
{\it Gaia} (\url{https://www.cosmos.esa.int/gaia}), processed by the {\it Gaia}
Data Processing and Analysis Consortium (DPAC,
\url{https://www.cosmos.esa.int/web/gaia/dpac/consortium}). Funding for the DPAC
has been provided by national institutions, in particular the institutions
participating in the {\it Gaia} Multilateral Agreement.
This publication makes use of data products from the Two Micron All Sky Survey, which is a joint project of the University of Massachusetts and the Infrared Processing and Analysis Center/California Institute of Technology, funded by the National Aeronautics and Space Administration and the National Science Foundation.
This publication makes use of data products from the Wide-field Infrared Survey Explorer, which is a joint project of the University of California, Los Angeles, and the Jet Propulsion Laboratory/California Institute of Technology, funded by the National Aeronautics and Space Administration.
Funding for the Sloan Digital Sky 
Survey IV has been provided by the 
Alfred P. Sloan Foundation, the U.S. 
Department of Energy Office of 
Science, and the Participating 
Institutions. SDSS-IV acknowledges support and 
resources from the Center for High 
Performance Computing  at the 
University of Utah. The SDSS 
website is www.sdss.org.
SDSS-IV is managed by the 
Astrophysical Research Consortium 
for the Participating Institutions 
of the SDSS Collaboration including 
the Brazilian Participation Group, 
the Carnegie Institution for Science, 
Carnegie Mellon University, Center for 
Astrophysics | Harvard \& 
Smithsonian, the Chilean Participation 
Group, the French Participation Group, 
Instituto de Astrof\'isica de 
Canarias, The Johns Hopkins 
University, Kavli Institute for the 
Physics and Mathematics of the 
Universe (IPMU) / University of 
Tokyo, the Korean Participation Group, 
Lawrence Berkeley National Laboratory, 
Leibniz Institut f\"ur Astrophysik 
Potsdam (AIP),  Max-Planck-Institut 
f\"ur Astronomie (MPIA Heidelberg), 
Max-Planck-Institut f\"ur 
Astrophysik (MPA Garching), 
Max-Planck-Institut f\"ur 
Extraterrestrische Physik (MPE), 
National Astronomical Observatories of 
China, New Mexico State University, 
New York University, University of 
Notre Dame, Observat\'ario 
Nacional / MCTI, The Ohio State 
University, Pennsylvania State 
University, Shanghai 
Astronomical Observatory, United 
Kingdom Participation Group, 
Universidad Nacional Aut\'onoma 
de M\'exico, University of Arizona, 
University of Colorado Boulder, 
University of Oxford, University of 
Portsmouth, University of Utah, 
University of Virginia, University 
of Washington, University of 
Wisconsin, Vanderbilt University, 
and Yale University.
Guoshoujing Telescope (the Large Sky Area Multi-Object Fiber Spectroscopic Telescope LAMOST) is a National Major Scientific Project built by the Chinese Academy of Sciences. Funding for the project has been provided by the National Development and Reform Commission. LAMOST is operated and managed by the National Astronomical Observatories, Chinese Academy of Sciences.

\section*{Data Availability}

The datasets used in this article were derived from sources in the public domain: Gaia EDR3, https://gea.esac.esa.int/archive/; UnWISE, https://catalog.unwise.me/catalogs.html; 2MASS, https://irsa.ipac.caltech.edu/Missions/2mass.html; LAMOST DR6, http://dr6.lamost.org/catalogue; SDSS DR17/APOGEE, https://skyserver.sdss.org/dr17.




\bibliographystyle{mnras}
\bibliography{Main_bib} 

\begin{thebibliography}{}
\makeatletter
\relax
\def\mn@urlcharsother{\let\do\@makeother \do\$\do\&\do\#\do\^\do\_\do\%\do\~}
\def\mn@doi{\begingroup\mn@urlcharsother \@ifnextchar [ {\mn@doi@}
  {\mn@doi@[]}}
\def\mn@doi@[#1]#2{\def\@tempa{#1}\ifx\@tempa\@empty \href
  {http://dx.doi.org/#2} {doi:#2}\else \href {http://dx.doi.org/#2} {#1}\fi
  \endgroup}
\def\mn@eprint#1#2{\mn@eprint@#1:#2::\@nil}
\def\mn@eprint@arXiv#1{\href {http://arxiv.org/abs/#1} {{\tt arXiv:#1}}}
\def\mn@eprint@dblp#1{\href {http://dblp.uni-trier.de/rec/bibtex/#1.xml}
  {dblp:#1}}
\def\mn@eprint@#1:#2:#3:#4\@nil{\def\@tempa {#1}\def\@tempb {#2}\def\@tempc
  {#3}\ifx \@tempc \@empty \let \@tempc \@tempb \let \@tempb \@tempa \fi \ifx
  \@tempb \@empty \def\@tempb {arXiv}\fi \@ifundefined
  {mn@eprint@\@tempb}{\@tempb:\@tempc}{\expandafter \expandafter \csname
  mn@eprint@\@tempb\endcsname \expandafter{\@tempc}}}

\bibitem[\protect\citeauthoryear{{Abbott} et~al.,}{{Abbott} et~al.}{2021}]{DES}
{Abbott} T.~M.~C.,  et~al., 2021, \mn@doi [\apjs] {10.3847/1538-4365/ac00b3},
  \href {https://ui.adsabs.harvard.edu/abs/2021ApJS..255...20A} {255, 20}

\bibitem[\protect\citeauthoryear{{Ahumada} et~al.,}{{Ahumada}
  et~al.}{2020}]{Ahumada2020}
{Ahumada} R.,  et~al., 2020, \mn@doi [\apjs] {10.3847/1538-4365/ab929e}, \href
  {https://ui.adsabs.harvard.edu/abs/2020ApJS..249....3A} {249, 3}

\bibitem[\protect\citeauthoryear{{Aihara} et~al.,}{{Aihara}
  et~al.}{2011}]{SDSS}
{Aihara} H.,  et~al., 2011, \mn@doi [\apjs] {10.1088/0067-0049/193/2/29}, \href
  {https://ui.adsabs.harvard.edu/abs/2011ApJS..193...29A} {193, 29}

\bibitem[\protect\citeauthoryear{{Anders} et~al.,}{{Anders}
  et~al.}{2022}]{Anders2022}
{Anders} F.,  et~al., 2022, \mn@doi [\aap] {10.1051/0004-6361/202142369}, \href
  {https://ui.adsabs.harvard.edu/abs/2022A&A...658A..91A} {658, A91}

\bibitem[\protect\citeauthoryear{{Anguiano} et~al.,}{{Anguiano}
  et~al.}{2018}]{Anguiano2018}
{Anguiano} B.,  et~al., 2018, \mn@doi [\aap] {10.1051/0004-6361/201833387},
  \href {https://ui.adsabs.harvard.edu/abs/2018A&A...620A..76A} {620, A76}

\bibitem[\protect\citeauthoryear{{Arenou} \& {Luri}}{{Arenou} \&
  {Luri}}{1999}]{Arenou1999}
{Arenou} F.,  {Luri} X.,  1999, in {Egret} D.,  {Heck} A.,  eds,  Astronomical
  Society of the Pacific Conference Series Vol. 167, Harmonizing Cosmic
  Distance Scales in a Post-HIPPARCOS Era. pp 13--32 (\mn@eprint {arXiv}
  {astro-ph/9812094})

\bibitem[\protect\citeauthoryear{{Arentsen} et~al.,}{{Arentsen}
  et~al.}{2020}]{PIGS}
{Arentsen} A.,  et~al., 2020, \mn@doi [\mnras] {10.1093/mnrasl/slz156}, \href
  {https://ui.adsabs.harvard.edu/abs/2020MNRAS.491L..11A} {491, L11}

\bibitem[\protect\citeauthoryear{{Babusiaux} et~al.,}{{Babusiaux}
  et~al.}{2010}]{Babusiaux2010}
{Babusiaux} C.,  et~al., 2010, \mn@doi [\aap] {10.1051/0004-6361/201014353},
  \href {https://ui.adsabs.harvard.edu/abs/2010A&A...519A..77B} {519, A77}

\bibitem[\protect\citeauthoryear{{Babusiaux} et~al.,}{{Babusiaux}
  et~al.}{2014}]{Babusiaux2014}
{Babusiaux} C.,  et~al., 2014, \mn@doi [\aap] {10.1051/0004-6361/201323044},
  \href {https://ui.adsabs.harvard.edu/abs/2014A&A...563A..15B} {563, A15}

\bibitem[\protect\citeauthoryear{{Bailer-Jones}}{{Bailer-Jones}}{2015}]{BailerJones2015}
{Bailer-Jones} C. A.~L.,  2015, \mn@doi [\pasp] {10.1086/683116}, \href
  {https://ui.adsabs.harvard.edu/abs/2015PASP..127..994B} {127, 994}

\bibitem[\protect\citeauthoryear{Bailer-Jones, Rybizki, Fouesneau, Demleitner
  \& Andrae}{Bailer-Jones et~al.}{2021}]{BailerJones2021}
Bailer-Jones C. A.~L.,  Rybizki J.,  Fouesneau M.,  Demleitner M.,   Andrae R.,
   2021, \mn@doi [The Astronomical Journal] {10.3847/1538-3881/abd806}, 161,
  147

\bibitem[\protect\citeauthoryear{{Barbuy}, {Chiappini}  \& {Gerhard}}{{Barbuy}
  et~al.}{2018}]{Barbuy2018}
{Barbuy} B.,  {Chiappini} C.,   {Gerhard} O.,  2018, \mn@doi [\araa]
  {10.1146/annurev-astro-081817-051826}, \href
  {https://ui.adsabs.harvard.edu/abs/2018ARA&A..56..223B} {56, 223}

\bibitem[\protect\citeauthoryear{{Belokurov}, {Erkal}, {Evans}, {Koposov}  \&
  {Deason}}{{Belokurov} et~al.}{2018}]{Belokurov2018}
{Belokurov} V.,  {Erkal} D.,  {Evans} N.~W.,  {Koposov} S.~E.,   {Deason}
  A.~J.,  2018, \mn@doi [\mnras] {10.1093/mnras/sty982}, \href
  {https://ui.adsabs.harvard.edu/abs/2018MNRAS.478..611B} {478, 611}

\bibitem[\protect\citeauthoryear{{Bianchi}, {Shiao}  \& {Thilker}}{{Bianchi}
  et~al.}{2017}]{GALEX}
{Bianchi} L.,  {Shiao} B.,   {Thilker} D.,  2017, \mn@doi [\apjs]
  {10.3847/1538-4365/aa7053}, \href
  {https://ui.adsabs.harvard.edu/abs/2017ApJS..230...24B} {230, 24}

\bibitem[\protect\citeauthoryear{{Bland-Hawthorn} \&
  {Gerhard}}{{Bland-Hawthorn} \& {Gerhard}}{2016}]{BHG}
{Bland-Hawthorn} J.,  {Gerhard} O.,  2016, \mn@doi [\araa]
  {10.1146/annurev-astro-081915-023441}, \href
  {https://ui.adsabs.harvard.edu/abs/2016ARA&A..54..529B} {54, 529}

\bibitem[\protect\citeauthoryear{{Buder} et~al.,}{{Buder}
  et~al.}{2019}]{Buder2019}
{Buder} S.,  et~al., 2019, \mn@doi [\aap] {10.1051/0004-6361/201833218}, \href
  {https://ui.adsabs.harvard.edu/abs/2019A&A...624A..19B} {624, A19}

\bibitem[\protect\citeauthoryear{{Buder} et~al.,}{{Buder}
  et~al.}{2021}]{Buder2020}
{Buder} S.,  et~al., 2021, \mn@doi [\mnras] {10.1093/mnras/stab1242}, \href
  {https://ui.adsabs.harvard.edu/abs/2021MNRAS.506..150B} {506, 150}

\bibitem[\protect\citeauthoryear{{Casey}, {Kennedy}, {Hartle}  \&
  {Schlaufman}}{{Casey} et~al.}{2018}]{Casey2018}
{Casey} A.~R.,  {Kennedy} G.~M.,  {Hartle} T.~R.,   {Schlaufman} K.~C.,  2018,
  \mn@doi [\mnras] {10.1093/mnras/sty1208}, \href
  {https://ui.adsabs.harvard.edu/abs/2018MNRAS.478.2812C} {478, 2812}

\bibitem[\protect\citeauthoryear{{Chambers} et~al.,}{{Chambers}
  et~al.}{2016}]{PanSTARRS}
{Chambers} K.~C.,  et~al., 2016, arXiv e-prints, \href
  {https://ui.adsabs.harvard.edu/abs/2016arXiv161205560C} {p. arXiv:1612.05560}

\bibitem[\protect\citeauthoryear{{Chawla}, {Bowyer}, {Hall}  \&
  {Kegelmeyer}}{{Chawla} et~al.}{2011}]{Chawla2011}
{Chawla} N.~V.,  {Bowyer} K.~W.,  {Hall} L.~O.,   {Kegelmeyer} W.~P.,  2011,
  arXiv e-prints, \href {https://ui.adsabs.harvard.edu/abs/2011arXiv1106.1813C}
  {p. arXiv:1106.1813}

\bibitem[\protect\citeauthoryear{{Chiti}, {Mardini}, {Frebel}  \&
  {Daniel}}{{Chiti} et~al.}{2021}]{Chiti2021}
{Chiti} A.,  {Mardini} M.~K.,  {Frebel} A.,   {Daniel} T.,  2021, \mn@doi
  [\apjl] {10.3847/2041-8213/abd629}, \href
  {https://ui.adsabs.harvard.edu/abs/2021ApJ...911L..23C} {911, L23}

\bibitem[\protect\citeauthoryear{{Cirasuolo} et~al.,}{{Cirasuolo}
  et~al.}{2014}]{MOONS}
{Cirasuolo} M.,  et~al., 2014, in {Ramsay} S.~K.,  {McLean} I.~S.,   {Takami}
  H.,  eds,  Society of Photo-Optical Instrumentation Engineers (SPIE)
  Conference Series Vol. 9147, Ground-based and Airborne Instrumentation for
  Astronomy V. p. 91470N, \mn@doi{10.1117/12.2056012}

\bibitem[\protect\citeauthoryear{Cui et~al.,}{Cui et~al.}{2012}]{Cui2012}
Cui X.-Q.,  et~al., 2012, \mn@doi [Research in Astronomy and Astrophysics]
  {10.1088/1674-4527/12/9/003}, 12, 1197

\bibitem[\protect\citeauthoryear{{DESI Collaboration} et~al.,}{{DESI
  Collaboration} et~al.}{2016}]{DESI}
{DESI Collaboration} et~al., 2016, arXiv e-prints, \href
  {https://ui.adsabs.harvard.edu/abs/2016arXiv161100036D} {p. arXiv:1611.00036}

\bibitem[\protect\citeauthoryear{{Dalton} et~al.,}{{Dalton}
  et~al.}{2012}]{Dalton2012}
{Dalton} G.,  et~al., 2012, in {McLean} I.~S.,  {Ramsay} S.~K.,   {Takami} H.,
  eds,  Society of Photo-Optical Instrumentation Engineers (SPIE) Conference
  Series Vol. 8446, Ground-based and Airborne Instrumentation for Astronomy IV.
  p. 84460P, \mn@doi{10.1117/12.925950}

\bibitem[\protect\citeauthoryear{{Dalton} et~al.,}{{Dalton}
  et~al.}{2014}]{WEAVE}
{Dalton} G.,  et~al., 2014, in {Ramsay} S.~K.,  {McLean} I.~S.,   {Takami} H.,
  eds,  Society of Photo-Optical Instrumentation Engineers (SPIE) Conference
  Series Vol. 9147, Ground-based and Airborne Instrumentation for Astronomy V.
  p. 91470L (\mn@eprint {arXiv} {1412.0843}), \mn@doi{10.1117/12.2055132}

\bibitem[\protect\citeauthoryear{{Debattista}, {Ness}, {Gonzalez}, {Freeman},
  {Zoccali}  \& {Minniti}}{{Debattista} et~al.}{2017}]{Debattista2017}
{Debattista} V.~P.,  {Ness} M.,  {Gonzalez} O.~A.,  {Freeman} K.,  {Zoccali}
  M.,   {Minniti} D.,  2017, \mn@doi [\mnras] {10.1093/mnras/stx947}, \href
  {https://ui.adsabs.harvard.edu/abs/2017MNRAS.469.1587D} {469, 1587}

\bibitem[\protect\citeauthoryear{{Gadotti} et~al.,}{{Gadotti}
  et~al.}{2019}]{Gadotti2019}
{Gadotti} D.~A.,  et~al., 2019, \mn@doi [\mnras] {10.1093/mnras/sty2666}, \href
  {https://ui.adsabs.harvard.edu/abs/2019MNRAS.482..506G} {482, 506}

\bibitem[\protect\citeauthoryear{{Gaia Collaboration} et~al.,}{{Gaia
  Collaboration} et~al.}{2016}]{Gaia2016}
{Gaia Collaboration} et~al., 2016, \mn@doi [\aap]
  {10.1051/0004-6361/201629272}, \href
  {https://ui.adsabs.harvard.edu/abs/2016A&A...595A...1G} {595, A1}

\bibitem[\protect\citeauthoryear{{Gaia Collaboration} et~al.,}{{Gaia
  Collaboration} et~al.}{2021}]{Gaia2021}
{Gaia Collaboration} et~al., 2021, \mn@doi [\aap]
  {10.1051/0004-6361/202039657}, \href
  {https://ui.adsabs.harvard.edu/abs/2021A&A...649A...1G} {649, A1}

\bibitem[\protect\citeauthoryear{{Gal} \& {Ghahramani}}{{Gal} \&
  {Ghahramani}}{2015}]{Gal2015}
{Gal} Y.,  {Ghahramani} Z.,  2015, arXiv e-prints, \href
  {https://ui.adsabs.harvard.edu/abs/2015arXiv150602142G} {p. arXiv:1506.02142}

\bibitem[\protect\citeauthoryear{{Gilmore} et~al.,}{{Gilmore}
  et~al.}{2012}]{GaiaESO}
{Gilmore} G.,  et~al., 2012, The Messenger, \href
  {https://ui.adsabs.harvard.edu/abs/2012Msngr.147...25G} {147, 25}

\bibitem[\protect\citeauthoryear{{Gonzalez} \& {Gadotti}}{{Gonzalez} \&
  {Gadotti}}{2016}]{Gonzalez2016}
{Gonzalez} O.~A.,  {Gadotti} D.,  2016, in {Laurikainen} E.,  {Peletier} R.,
  {Gadotti} D.,  eds,  Astrophysics and Space Science Library Vol. 418,
  Galactic Bulges. p.~199 (\mn@eprint {arXiv} {1503.07252}),
  \mn@doi{10.1007/978-3-319-19378-6\_9}

\bibitem[\protect\citeauthoryear{Grady, Belokurov  \& Evans}{Grady
  et~al.}{2021}]{Grady2021}
Grady J.,  Belokurov V.,   Evans N.~W.,  2021, \mn@doi [The Astrophysical
  Journal] {10.3847/1538-4357/abd4e4}, 909, 150

\bibitem[\protect\citeauthoryear{{Gustafsson}, {Edvardsson}, {Eriksson},
  {J{\o}rgensen}, {Nordlund}  \& {Plez}}{{Gustafsson} et~al.}{2008}]{MARCS}
{Gustafsson} B.,  {Edvardsson} B.,  {Eriksson} K.,  {J{\o}rgensen} U.~G.,
  {Nordlund} {\AA}.,   {Plez} B.,  2008, \mn@doi [\aap]
  {10.1051/0004-6361:200809724}, \href
  {http://adsabs.harvard.edu/abs/2008A%26A...486..951G} {486, 951}

\bibitem[\protect\citeauthoryear{{Hinton}, {Srivastava}, {Krizhevsky},
  {Sutskever}  \& {Salakhutdinov}}{{Hinton} et~al.}{2012}]{Hinton2012}
{Hinton} G.~E.,  {Srivastava} N.,  {Krizhevsky} A.,  {Sutskever} I.,
  {Salakhutdinov} R.~R.,  2012, arXiv e-prints, \href
  {https://ui.adsabs.harvard.edu/abs/2012arXiv1207.0580H} {p. arXiv:1207.0580}

\bibitem[\protect\citeauthoryear{{Hogg}, {Eilers}  \& {Rix}}{{Hogg}
  et~al.}{2019}]{Hogg2019}
{Hogg} D.~W.,  {Eilers} A.-C.,   {Rix} H.-W.,  2019, \mn@doi [\aj]
  {10.3847/1538-3881/ab398c}, \href
  {https://ui.adsabs.harvard.edu/abs/2019AJ....158..147H} {158, 147}

\bibitem[\protect\citeauthoryear{{Huang} et~al.,}{{Huang}
  et~al.}{2022}]{Huang2021}
{Huang} Y.,  et~al., 2022, \mn@doi [\apj] {10.3847/1538-4357/ac21cb}, \href
  {https://ui.adsabs.harvard.edu/abs/2022ApJ...925..164H} {925, 164}

\bibitem[\protect\citeauthoryear{{Ivezi{\'c}} et~al.,}{{Ivezi{\'c}}
  et~al.}{2008}]{Ivezi2008}
{Ivezi{\'c}} {\v{Z}}.,  et~al., 2008, \mn@doi [\apj] {10.1086/589678}, \href
  {https://ui.adsabs.harvard.edu/abs/2008ApJ...684..287I} {684, 287}

\bibitem[\protect\citeauthoryear{{Keller} et~al.,}{{Keller}
  et~al.}{2007}]{SkyMapper_design}
{Keller} S.~C.,  et~al., 2007, \mn@doi [\pasa] {10.1071/AS07001}, \href
  {https://ui.adsabs.harvard.edu/abs/2007PASA...24....1K} {24, 1}

\bibitem[\protect\citeauthoryear{{Kingma} \& {Welling}}{{Kingma} \&
  {Welling}}{2019}]{Kingma2019}
{Kingma} D.~P.,  {Welling} M.,  2019, arXiv e-prints, \href
  {https://ui.adsabs.harvard.edu/abs/2019arXiv190602691K} {p. arXiv:1906.02691}

\bibitem[\protect\citeauthoryear{{Kollmeier} et~al.,}{{Kollmeier}
  et~al.}{2017}]{MWM}
{Kollmeier} J.~A.,  et~al., 2017, arXiv e-prints, \href
  {https://ui.adsabs.harvard.edu/abs/2017arXiv171103234K} {p. arXiv:1711.03234}

\bibitem[\protect\citeauthoryear{{Koposov}, {Belokurov}, {Zucker}, {Lewis},
  {Ibata}, {Olszewski}, {L{\'o}pez-S{\'a}nchez}  \& {Hyde}}{{Koposov}
  et~al.}{2015}]{Koposov2015}
{Koposov} S.~E.,  {Belokurov} V.,  {Zucker} D.~B.,  {Lewis} G.~F.,  {Ibata}
  R.~A.,  {Olszewski} E.~W.,  {L{\'o}pez-S{\'a}nchez} {\'A}.~R.,   {Hyde}
  E.~A.,  2015, \mn@doi [\mnras] {10.1093/mnras/stu2263}, \href
  {http://adsabs.harvard.edu/abs/2015MNRAS.446.3110K} {446, 3110}

\bibitem[\protect\citeauthoryear{{Leung} \& {Bovy}}{{Leung} \&
  {Bovy}}{2019a}]{Leung2019a}
{Leung} H.~W.,  {Bovy} J.,  2019a, \mn@doi [\mnras] {10.1093/mnras/sty3217},
  \href {https://ui.adsabs.harvard.edu/abs/2019MNRAS.483.3255L} {483, 3255}

\bibitem[\protect\citeauthoryear{Leung \& Bovy}{Leung \&
  Bovy}{2019b}]{Leung2019b}
Leung H.~W.,  Bovy J.,  2019b, \mn@doi [Monthly Notices of the Royal
  Astronomical Society] {10.1093/mnras/stz2245}, 489, 2079

\bibitem[\protect\citeauthoryear{Li et~al.,}{Li et~al.}{2016}]{Li2016}
Li J.,  et~al., 2016, \mn@doi [The Astrophysical Journal]
  {10.3847/0004-637x/823/1/59}, 823, 59

\bibitem[\protect\citeauthoryear{{Lin}, {Casagrande}  \& {Asplund}}{{Lin}
  et~al.}{2022}]{Lin2022}
{Lin} J.,  {Casagrande} L.,   {Asplund} M.,  2022, \mn@doi [\mnras]
  {10.1093/mnras/stab3326}, \href
  {https://ui.adsabs.harvard.edu/abs/2022MNRAS.510..433L} {510, 433}

\bibitem[\protect\citeauthoryear{{Lindegren} et~al.,}{{Lindegren}
  et~al.}{2021}]{Lindegren2021}
{Lindegren} L.,  et~al., 2021, \mn@doi [\aap] {10.1051/0004-6361/202039653},
  \href {https://ui.adsabs.harvard.edu/abs/2021A&A...649A...4L} {649, A4}

\bibitem[\protect\citeauthoryear{{Majewski}, {Skrutskie}, {Weinberg}  \&
  {Ostheimer}}{{Majewski} et~al.}{2003}]{Majewski2003}
{Majewski} S.~R.,  {Skrutskie} M.~F.,  {Weinberg} M.~D.,   {Ostheimer} J.~C.,
  2003, \mn@doi [\apj] {10.1086/379504}, \href
  {https://ui.adsabs.harvard.edu/abs/2003ApJ...599.1082M} {599, 1082}

\bibitem[\protect\citeauthoryear{{Majewski}, {Zasowski}  \&
  {Nidever}}{{Majewski} et~al.}{2011}]{Majewski2011}
{Majewski} S.~R.,  {Zasowski} G.,   {Nidever} D.~L.,  2011, \mn@doi [\apj]
  {10.1088/0004-637X/739/1/25}, \href
  {https://ui.adsabs.harvard.edu/abs/2011ApJ...739...25M} {739, 25}

\bibitem[\protect\citeauthoryear{{Majewski} et~al.,}{{Majewski}
  et~al.}{2017}]{Majewski2017}
{Majewski} S.~R.,  et~al., 2017, \mn@doi [\aj] {10.3847/1538-3881/aa784d},
  \href {https://ui.adsabs.harvard.edu/abs/2017AJ....154...94M} {154, 94}

\bibitem[\protect\citeauthoryear{{Minniti}, {Olszewski}, {Liebert}, {White},
  {Hill}  \& {Irwin}}{{Minniti} et~al.}{1995}]{Minniti1995}
{Minniti} D.,  {Olszewski} E.~W.,  {Liebert} J.,  {White} S. D.~M.,  {Hill}
  J.~M.,   {Irwin} M.~J.,  1995, \mn@doi [\mnras] {10.1093/mnras/277.4.1293},
  \href {https://ui.adsabs.harvard.edu/abs/1995MNRAS.277.1293M} {277, 1293}

\bibitem[\protect\citeauthoryear{{Myeong}, {Vasiliev}, {Iorio}, {Evans}  \&
  {Belokurov}}{{Myeong} et~al.}{2019}]{Myeong2019}
{Myeong} G.~C.,  {Vasiliev} E.,  {Iorio} G.,  {Evans} N.~W.,   {Belokurov} V.,
  2019, \mn@doi [\mnras] {10.1093/mnras/stz1770}, \href
  {https://ui.adsabs.harvard.edu/abs/2019MNRAS.488.1235M} {488, 1235}

\bibitem[\protect\citeauthoryear{{Ness} \& {Freeman}}{{Ness} \&
  {Freeman}}{2016}]{Ness2016}
{Ness} M.,  {Freeman} K.,  2016, \mn@doi [\pasa] {10.1017/pasa.2015.51}, \href
  {https://ui.adsabs.harvard.edu/abs/2016PASA...33...22N} {33, e022}

\bibitem[\protect\citeauthoryear{{Ness} et~al.,}{{Ness}
  et~al.}{2013}]{Ness2013}
{Ness} M.,  et~al., 2013, \mn@doi [\mnras] {10.1093/mnras/sts629}, \href
  {https://ui.adsabs.harvard.edu/abs/2013MNRAS.430..836N} {430, 836}

\bibitem[\protect\citeauthoryear{{Paszke} et~al.,}{{Paszke}
  et~al.}{2019}]{Paszke2019}
{Paszke} A.,  et~al., 2019, arXiv e-prints, \href
  {https://ui.adsabs.harvard.edu/abs/2019arXiv191201703P} {p. arXiv:1912.01703}

\bibitem[\protect\citeauthoryear{{Rich}, {Origlia}  \& {Valenti}}{{Rich}
  et~al.}{2012}]{Rich2012}
{Rich} R.~M.,  {Origlia} L.,   {Valenti} E.,  2012, \mn@doi [\apj]
  {10.1088/0004-637X/746/1/59}, \href
  {https://ui.adsabs.harvard.edu/abs/2012ApJ...746...59R} {746, 59}

\bibitem[\protect\citeauthoryear{{Riello} et~al.,}{{Riello}
  et~al.}{2021}]{Riello2021}
{Riello} M.,  et~al., 2021, \mn@doi [\aap] {10.1051/0004-6361/202039587}, \href
  {https://ui.adsabs.harvard.edu/abs/2021A&A...649A...3R} {649, A3}

\bibitem[\protect\citeauthoryear{{Saglia}, {Opitsch}, {Fabricius}, {Bender},
  {Bla{\~n}a}  \& {Gerhard}}{{Saglia} et~al.}{2018}]{Saglia2018}
{Saglia} R.~P.,  {Opitsch} M.,  {Fabricius} M.~H.,  {Bender} R.,  {Bla{\~n}a}
  M.,   {Gerhard} O.,  2018, \mn@doi [\aap] {10.1051/0004-6361/201732517},
  \href {https://ui.adsabs.harvard.edu/abs/2018A&A...618A.156S} {618, A156}

\bibitem[\protect\citeauthoryear{{Schlafly}, {Meisner}  \& {Green}}{{Schlafly}
  et~al.}{2019}]{Schlafly2019}
{Schlafly} E.~F.,  {Meisner} A.~M.,   {Green} G.~M.,  2019, \mn@doi [\apjs]
  {10.3847/1538-4365/aafbea}, \href
  {https://ui.adsabs.harvard.edu/abs/2019ApJS..240...30S} {240, 30}

\bibitem[\protect\citeauthoryear{{Schlaufman} \& {Casey}}{{Schlaufman} \&
  {Casey}}{2014}]{Schlaufman2014}
{Schlaufman} K.~C.,  {Casey} A.~R.,  2014, \mn@doi [\apj]
  {10.1088/0004-637X/797/1/13}, \href
  {https://ui.adsabs.harvard.edu/abs/2014ApJ...797...13S} {797, 13}

\bibitem[\protect\citeauthoryear{{Seabroke} et~al.,}{{Seabroke}
  et~al.}{2021}]{Seabroke2021}
{Seabroke} G.,  et~al., 2021, arXiv e-prints, \href
  {https://ui.adsabs.harvard.edu/abs/2021arXiv210802796S} {p. arXiv:2108.02796}

\bibitem[\protect\citeauthoryear{{Skrutskie} et~al.,}{{Skrutskie}
  et~al.}{2006}]{Skrutskie2006}
{Skrutskie} M.~F.,  et~al., 2006, \mn@doi [\aj] {10.1086/498708}, \href
  {https://ui.adsabs.harvard.edu/abs/2006AJ....131.1163S} {131, 1163}

\bibitem[\protect\citeauthoryear{Stan Dev~Team}{Stan
  Dev~Team}{2021}]{standevteam2021}
Stan Dev~Team .,  2021, cmdstanpy, \url {https://pypi.org/project/cmdstanpy/}

\bibitem[\protect\citeauthoryear{{Starkenburg} et~al.,}{{Starkenburg}
  et~al.}{2017}]{PRISTINE}
{Starkenburg} E.,  et~al., 2017, \mn@doi [\mnras] {10.1093/mnras/stx1068},
  \href {https://ui.adsabs.harvard.edu/abs/2017MNRAS.471.2587S} {471, 2587}

\bibitem[\protect\citeauthoryear{{Steinmetz} et~al.,}{{Steinmetz}
  et~al.}{2020}]{Steinmetz2020}
{Steinmetz} M.,  et~al., 2020, \mn@doi [\aj] {10.3847/1538-3881/ab9ab9}, \href
  {https://ui.adsabs.harvard.edu/abs/2020AJ....160...82S} {160, 82}

\bibitem[\protect\citeauthoryear{{Thomas} et~al.,}{{Thomas}
  et~al.}{2019}]{Thomas2019}
{Thomas} G.~F.,  et~al., 2019, \mn@doi [\apj] {10.3847/1538-4357/ab4a77}, \href
  {https://ui.adsabs.harvard.edu/abs/2019ApJ...886...10T} {886, 10}

\bibitem[\protect\citeauthoryear{{Wallerstein}}{{Wallerstein}}{1962}]{Wallerstein1962}
{Wallerstein} G.,  1962, \mn@doi [\apjs] {10.1086/190067}, \href
  {https://ui.adsabs.harvard.edu/abs/1962ApJS....6..407W} {6, 407}

\bibitem[\protect\citeauthoryear{{Wang} \& {Chen}}{{Wang} \&
  {Chen}}{2019}]{Wang2019}
{Wang} S.,  {Chen} X.,  2019, \mn@doi [\apj] {10.3847/1538-4357/ab1c61}, \href
  {https://ui.adsabs.harvard.edu/abs/2019ApJ...877..116W} {877, 116}

\bibitem[\protect\citeauthoryear{{Wegg}, {Rojas-Arriagada}, {Schultheis}  \&
  {Gerhard}}{{Wegg} et~al.}{2019}]{Wegg2019}
{Wegg} C.,  {Rojas-Arriagada} A.,  {Schultheis} M.,   {Gerhard} O.,  2019,
  \mn@doi [\aap] {10.1051/0004-6361/201936779}, \href
  {https://ui.adsabs.harvard.edu/abs/2019A&A...632A.121W} {632, A121}

\bibitem[\protect\citeauthoryear{{Wolf} et~al.,}{{Wolf}
  et~al.}{2018}]{SkyMapper}
{Wolf} C.,  et~al., 2018, \mn@doi [\pasa] {10.1017/pasa.2018.5}, \href
  {https://ui.adsabs.harvard.edu/abs/2018PASA...35...10W} {35, e010}

\bibitem[\protect\citeauthoryear{{Wright} et~al.,}{{Wright}
  et~al.}{2010}]{Wright2010}
{Wright} E.~L.,  et~al., 2010, \mn@doi [\aj] {10.1088/0004-6256/140/6/1868},
  \href {https://ui.adsabs.harvard.edu/abs/2010AJ....140.1868W} {140, 1868}

\bibitem[\protect\citeauthoryear{{Yanny} et~al.,}{{Yanny}
  et~al.}{2009}]{Yanny2009}
{Yanny} B.,  et~al., 2009, \mn@doi [\aj] {10.1088/0004-6256/137/5/4377}, \href
  {https://ui.adsabs.harvard.edu/abs/2009AJ....137.4377Y} {137, 4377}

\bibitem[\protect\citeauthoryear{{Zhao}, {Spergel}  \& {Rich}}{{Zhao}
  et~al.}{1994}]{Zhao1994}
{Zhao} H.,  {Spergel} D.~N.,   {Rich} R.~M.,  1994, \mn@doi [\aj]
  {10.1086/117227}, \href
  {https://ui.adsabs.harvard.edu/abs/1994AJ....108.2154Z} {108, 2154}

\bibitem[\protect\citeauthoryear{{Zoccali}, {Hill}, {Lecureur}, {Barbuy},
  {Renzini}, {Minniti}, {Gomez}  \& {Ortolani}}{{Zoccali}
  et~al.}{2008}]{Zoccali2008}
{Zoccali} M.,  {Hill} V.,  {Lecureur} A.,  {Barbuy} B.,  {Renzini} A.,
  {Minniti} D.,  {Gomez} A.,   {Ortolani} S.,  2008, VizieR Online Data
  Catalog, \href {https://ui.adsabs.harvard.edu/abs/2008yCat..34860177Z} {pp
  J/A+A/486/177}

\bibitem[\protect\citeauthoryear{{de Jong} et~al.,}{{de Jong}
  et~al.}{2019}]{Jong21019}
{de Jong} R.~S.,  et~al., 2019, \mn@doi [The Messenger]
  {10.18727/0722-6691/5117}, \href
  {https://ui.adsabs.harvard.edu/abs/2019Msngr.175....3D} {175, 3}

\makeatother
\end{thebibliography}



\appendix

\section{Neural Network Setup}
\label{appendix_NNsetup}

Here we describe the specific inputs and architecture we use to design our neural networks.

\subsection{Input Features}

We select 24 features to use as input for our NN: 16 colours, and 8 absolute magnitudes. Constructed from Gaia G, G$_{RP}$, \& G$_{BP}$, 2MASS J, H, \& K$_{s}$, and WISE W1 \& W2 photometric bands, and extinction corrected following the RJCE method \citep{Majewski2011}, we select the following colours:

(J-K$_{s}$), (J-H), (H-K$_{s}$), (W1-W2), (G$_{\mathrm{BP}}$-J), (G$_{\mathrm{BP}}$-H), (G$_{\mathrm{BP}}$-K$_{s}$), (G$_{\mathrm{BP}}$-W1), (G$_{\mathrm{BP}}$-W2), (G$_{\mathrm{RP}}$-K$_{s}$), (G$_{\mathrm{BP}}$-G$_{\mathrm{RP}}$), (G$_{\mathrm{BP}}$-G$_{\mathrm{G}}$), (G$_{\mathrm{G}}$-G$_{\mathrm{RP}}$), (J-W1), (J-W2), (H-W2).

We also include the following absolute magnitudes when noted, calculated with distances from our NN-enhanced approach:

BP, RP, G, W1, W2, J, H, and K$_{s}$.

\subsection{Network Architecture}

We build our network out of four main layers: an input layer, two hidden layers, and an output layer.

Out input layer accepts the 24 input features, and assigns each of them to a node in the network. We have two hidden layers, both with 80 nodes, which are fully interconnected between each other, the input layer, and the output layer. These hidden layers also have drop-out applied, with a weighting of 20\% for each pass (i.e.: one fifth of each hidden layer is `dropped' each run) of the network during training or prediction. Two layers of 80 hidden nodes were chosen as the result of manual tuning, where we found a large network with drop-out gave the best recovery of initial data while maintaining the network's confidence in it's predictions.

Our output layer contains only two nodes: the output node, where we return the `final' output; and an uncertainty node which recovers the network's certainty in it's prediction. This is explained fully in Section~\ref{NNsetup}.

\section{Photometric Metallicity Comparisons}
\label{appendix_fehcomp}

\begin{figure*}
	\centering
	\includegraphics[width=.48\textwidth, trim={0, 0, 30, 30}, clip]{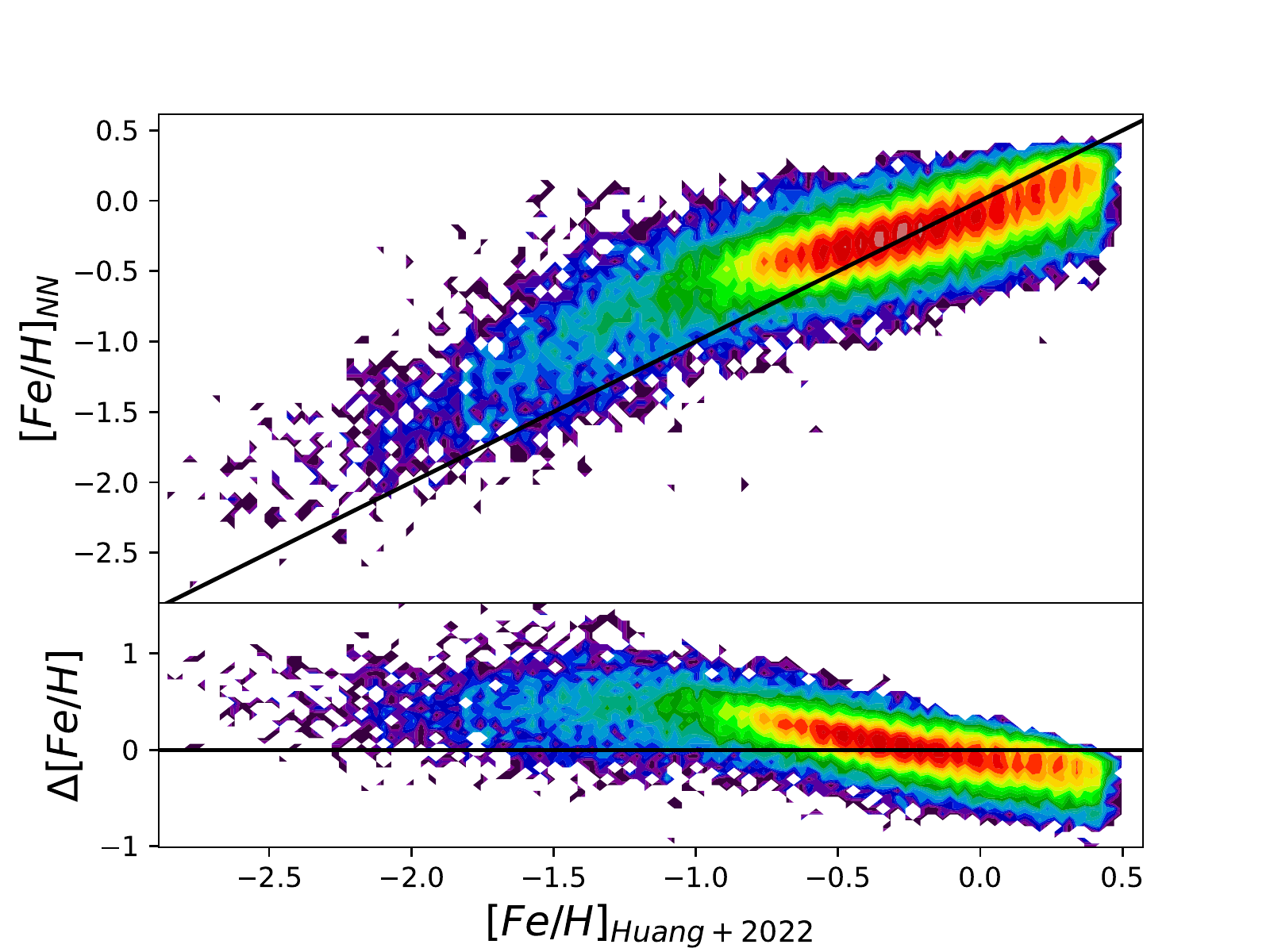}
	\includegraphics[width=.48\textwidth, trim={0, 0, 30, 30}, clip]{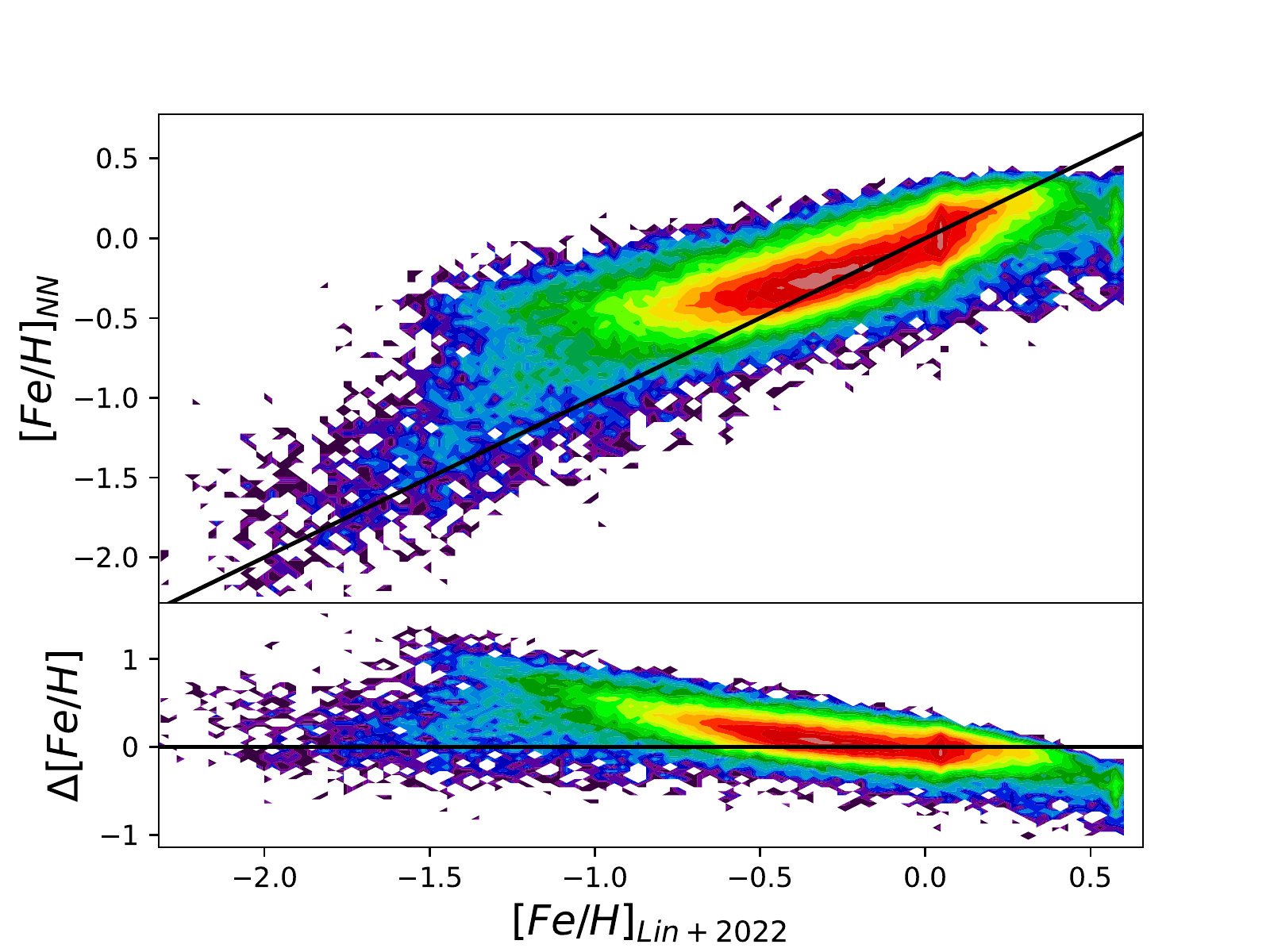}
    \caption{Plot of comparisons between our NN-estimated metallicities with those from \citet{Huang2021} (left) and \citet{Lin2022} (right).}
    \label{fig:feh_comp}
\end{figure*}

We compare the metallicities returned by our method to similar photometric-based techniques from \cite{Huang2021} and \cite{Lin2022}, as shown in Fig~\ref{fig:feh_comp}. Both papers use Skymapper $u$ and $v$ photometry with \cite{Lin2022} comparing to theoretical isochrones and \cite{Huang2021} using a data-driven approach deriving polynomial colour relations for the metallicities fitted to SDSS (APOGEE DR14 \& DR16) and LAMOST (DR7) data. We find a good correlation with these studies, especially at low metallicities ($\mathrm{[Fe/H]} < -1.5$). There is, however, a notable over-estimation in our metallicities visible in the \citet{Lin2022} comparison at $-1.5 < \mathrm{[Fe/H]} < -0.5$, where our method appears to predict a large portion of the sample with $\mathrm{[Fe/H]} \approx -0.5$. We note that the objects that make up this bias do tend to be stars with low $\log g$ values, suggesting this is may be a regime where the NN under-performs, possibly due to lack of training data. Alternatively, this bias could be due to discrepancies in the isochrones utilised by \citet{Lin2022} for cool stars.

\bsp	
\label{lastpage}
\end{document}